%% file: patbasis.tex
\documentclass[12pt]{article}

\pdfoutput=1 

\usepackage{hyperref} 

\usepackage{patbasis-macros} 

\usepackage{tikz} 

\usepackage{patbasis-title} 

\usepackage{patbasis-abstract} 

\usepackage{patbasis-intro} 

\usepackage{patbasis-summary} 

\usepackage{patbasis-figures} 

\usepackage{patbasis-definitions} 

\usepackage{patbasis-motivation} 

\usepackage{patbasis-discussion} 

\begin{document} 

\maketitle 

\begin{abstract}         

\myabstract

\end{abstract}

\newpage

\tableofcontents 

\newpage 

\section{Introduction\label{S:Introduction}}       

\myintro

\section{Summary of approach and ideas\label{S:Summary}}       


\subsection{background and terminology}

\subsubsection{circuit models}\label{S:CircuitModels}

Given a specific \emphdef{circuit model}
(set of \emphdef{primitive gates}, \emphdef{construction rules} for combining them into circuits,
and definition of nonnegative \emphdef{circuit cost} which is additive when circuits are serially composed),
the \emphdef{circuit complexity} of a function is defined as the minimum cost of any constructible circuit 
which computes it
(or infinity, if no constructible circuit computes it).

We don't require that a circuit model can construct circuits which compute every function, 
nor that it treat common simple operations (like interchanging the order of wires) as having zero cost (or even as always being permitted),
since we want the general framework to apply to intentionally limited models,
such as models which can construct only:
\begin{itemize}
\item
reversible functions ---
e.g. all primitive gates are small (reversible) \boldquote{S-boxes}\footnote{
    In this paper, by ``S-box'' we mean a reversible boolean function on a small number of bits
    (or a gate which computes such a function);
    by ``reversible'' we mean bijective (invertible and onto).
},
and wires can't be split or discarded
(typically 
this can construct 
any even permutation of the $2^n$ input vectors \cite{CG75}); or
\item
$\F_2$-linear functions --- e.g. wires {\it can} be split or discarded,
but only XOR gates (which do addition in $\F_2$) can be used.
\end{itemize}
But our main focus in this paper is on a conventional model for \boldquote{general boolean computation},
whose circuits can compute any boolean function (with any number of inputs and outputs), 
and which treats permuting or splitting wires as free of cost. (Such models are well-known,
and the choice among them has only a linear effect on circuit complexity \cite{Wil11}, 
so we have no need in this paper to pick a specific one.)

Without loss of generality, 
we assume each circuit model also specifies a \emphdef{generating set} $\mathcal{G}$
of \emphdef{generating circuits},
which are sufficient (when serially composed) to generate the same constructible functions without imposing a cost penalty ---
that is, the minimum cost of computing any function using a serial composition of generating circuits
is the same as the minimum over all constructible circuits in the model.\footnote{
Note that in spite of this assumption, there is no guarantee
that a generating circuit is the lowest-cost way to compute its own function ---
some composition of other generating circuits might in principle have lower cost.
}
(The set $\mathcal{G}$ will be infinite in models permitting arbitrarily high ``circuit widths'', 
but would typically have a simple structure;
see Section~\ref{S:StructureOfGeneratingCircuits}
for a conventional construction of $\mathcal{G}$.)

Without this assumption, 
it would be harder to analyze circuit cost when 
the model's construction rules include \boldquote{parallel composition},
i.e. placing non-inter\-acting circuits side by side
to make one larger circuit, combining their respective sets of inputs and outputs by disjoint union.
We don't want to require every model to simply add costs in that case (though typical models would do that); 
by making the model predefine an adequate set of generating circuits, 
which need only serial composition to be further combined, 
we avoid that issue.

This assumption is reasonable in practice --- in a typical model, the generating circuits would include
any circuit containing at most one primitive gate or trivial wire operation (like ``splitting a wire''\footnote{
    In most formalizations of circuits,
    all gates have exactly one output,
    which can be ``split'' into multiple wires
    by giving the gate a ``fanout'' greater than 1.
    In our formalization, this would be confusing
    (since gates are allowed to have multiple outputs with different values);
    for that and other reasons,
    we replace the concept of ``fanout''
    by the operation of ``splitting a wire'',
    which doesn't directly involve any gates
    (though it could alternatively be thought of 
    as a gate
    with one input and two outputs, all with equal values). 
    Similarly, by ``discarding a wire''
    we mean letting it ``end'' within the circuit;
    this could be thought of as a gate
    with one input and no outputs.
}),
alongside any number of non-interacting parallel wires (possibly permuted); then
any circuit could simply be {\it viewed} as a serial composition of generating circuits,
without affecting its cost.
Accordingly, from now on \boldquote{composition of circuits} always means
``serial composition'', which also composes the functions computed by those circuits.

\subsubsection{illustration of how a circuit is composed of generating circuits\label{S:illustrationofgeneratingcircuits}}


\noindent \textbf{Figure 1:} A simple multiplexer circuit, shown in a conventional form.

\smallskip
\smallskip
\smallskip
\centerline{\input pbcircuit-modeleg.tex} 

\smallskip
\smallskip
\smallskip
\smallskip
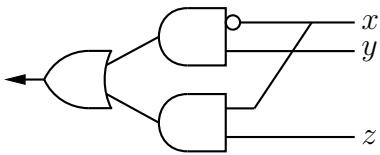
\noindent \textbf{Figure 2:} The same circuit as in Figure~1, 
shown as a composition of 6 generating circuits
(which have the ``conventional'' structure described in 
Section~\ref{S:StructureOfGeneratingCircuits}).

\smallskip
\smallskip
\smallskip
\centerline{\input pbcircuit-modeleg-comp.tex} 

The point of this circuit representation is that any function computed by a whole circuit
is a composition of functions computed by only a few types of generating circuit.
The generating circuits used here include 
3 gates, 2 wire permutations, and a ``splitter'',
in some cases with 1 or 2 ``adjoined wires'' 
(shown below the gate or splitter, 
since they correspond to their generating circuit's highest-numbered inputs and outputs, 
but logically running ``alongside'' or ``in parallel to'' the gate or splitter).

The ``widest'' generating circuits in this example 
have 4 inputs and 3 or 4 outputs: 
the AND gate with 2 adjoined wires, 
and the permutation of 4 wires. 
(The circuit as a whole has 3 inputs and 1 output, just as in Figure~1.)

\subsubsection{``complexity formula''}

By a \emphdef{complexity formula} (for a specific circuit model),
we mean a formula which can be applied to the truth table of an arbitrary function $f$,
which always produces a lower bound on its actual circuit complexity (for that circuit model).
The function computed by such a formula can be called a \emphdef{complexity measure}, 
or ``a kind of \emphdef{measured complexity}'' when we want to emphasize that it's only a lower bound, 
not necessarily a good one.
We are just as concerned with the formula itself (an expression of a certain form) 
as with the function it computes, 
since we need to ensure it's possible to prove lower bounds on its values for certain kinds of functions.

We informally call a complexity formula \emphdef{useful} if it produces reasonably good lower bounds,
at least for some functions of special interest and for most arbitrary functions.
(Note that most functions have near-maximal complexity. 
The requirement of being useful for most functions
might not matter for some applications,
but for an approach like this one
(which would prove its lower bound inductively over function composition),
we believe it's necessary,
for well-known reasons we'll discuss later.)

In this paper, by \emphdef{complexity} alone
we mean either circuit or measured complexity, according to context.
(When other kinds of complexity are needed, we'll name them explicitly.)

\subsection{general expectations}

We expect that for a given circuit model,
there may be more than one kind of useful measured complexity,
and probably no single formula (of a useful form) gives an exact measure of complexity for all functions.

Even a single example of a useful complexity formula might be {\it very} useful.
But given these expectations, and guessing there might be no canonical choice of useful complexity formula,
we're interested in techniques for coming up with families of (provably valid) complexity formulae
which are candidates for being useful. (This paper presents one such technique.)

\subsection{Basic idea: we want a sum of nonnegative terms...\label{S:BasicIdea}}

Now we turn to the basic ideas behind our proposed approach.
To start with,
we'll describe it ``from the outside in'', 
so we can discuss
how it might evade the natural proofs barrier
before anything else.
The remainder of Section~\ref{S:Summary} will fill in the details,
including what's still required to make the framework useful.
(The figures in Section~\ref{S:Figures}
illustrate how 
the proposed kind of formula would be computed 
for some simple examples, 
and compare the effects of several choices of pattern basis.
See also Section~\ref{S:Motivation} (Motivation)
for a presentation from a completely different direction.)

\bigskip
The proposed formula for a complexity lower bound is 
analogous to
``the log of the sum
of a very large number of nonnegative terms'' --- so many terms that computing the log of the sum directly
(from an arbitrary truth table)
takes too long to be
ruled out by the natural proofs barrier;
but with all terms nonnegative, so
lower bounds on the formula's value for specific functions
can in principle be proven, if enough is known about those functions to prove many specific terms are positive.
(The exact nature of the formula, and of the
``sum of terms'' it contains, 
will be discussed below.
See Section~\ref{S:naturalize} for a more technical discussion 
of why the natural proofs barrier seems unlikely to rule out the proposed kind of formula.)

One of our goals is to construct the formula so that,
for any function~$f$, 
most of the summed terms are zero (or low-valued enough to be insignificant),
and (for most~$f$)
the relatively few nonzero terms are ``difficult to find'' (in spite of their absolute number being large).
This means you can't even {\it estimate}~
$f$'s complexity by random probing of only a polynomial number of terms
(relative to the size of~$f$'s truth table),
since it's likely all the terms you probe will be zero\footnote{
    Except possibly for a few 
    ``non-discriminating'' 
    terms (which are each nonzero for {\it many} functions), which we can ignore here.
}~---
unless you know {\it which} terms to probe (since you know or guess something about~$f$'s structure),
or unless you probe all the ``simple'' terms and some of {\it those} are nonzero
(which is often true about special or interesting~$f$, but isn't true for most~$f$).
This is important, since otherwise a randomized polynomial algorithm could estimate measured complexity
well enough to distinguish low-complexity functions from random functions,
which would give us a ``useful, large, and constructive'' property (in the terminology of the natural proofs barrier)~---
that is, it would prove (subject to widely believed assumptions about pseudorandomness) 
that our approach can't be made to work.

\bigskip
As mentioned above, the terms in our formula's sum are always nonnegative~---
this lets you prove lower bounds on the sum (and thus on its log, and thus on $f$'s complexity)
if you know the ``location'' (i.e. the \emphdef{term index}, or position in the expression for the sum)
of enough positive terms --- which you might know,
if you understand the structure of~$f$, for example for explicit functions~$f$.
(Furthermore, if you don't need an {\it optimal} bound,
it's sufficient to lower-bound only {\it some} of the positive terms, which means
partial knowledge of~$f$'s structure or of~$f$ itself
might still allow you to prove a useful lower bound on its complexity.)

\subsection{... which are magnitudes of pattern match values ...}

The terms in the ``sum'' are the nonnegative \boldquote{magnitudes}
of \emphdef{pattern match values} (or \emphdef{pattern values} for short),
which are numbers (possibly negative or complex) in some field $K$
(probably $\C$ or $\R$, or possibly a finite field).
(The magnitude measure $M \colon K \to \R^{\ge 0}$ (on pattern values) could be any submultiplicative and subadditive function,
such as 0 for 0 and 1 for all other values, any submultiplicative norm over $K$, or other possibilities.
Generalizing $K$ to a perhaps-noncommutative ring might be worth considering, but is not further discussed here,
except to note that many of our definitions and proofs would work without change in that case.)

The intuitive interpretation is that each term measures the degree
to which $f$'s truth table ``matches a specific pattern'';
the pattern value (and therefore the term which measures its magnitude,
which we'll also call a \emphdef{pattern magnitude})
can be computed from the truth table given the term index.
(Note that for most patterns, the functions to compute their values would be very complex,
and whatever was special about the matching truth tables would not be something ``visible'' or ``obvious''.
It's no coincidence that the patterns we usually ``see'' or measure are the ``simple'' ones,
which are a small subset of all patterns, as discussed further below.)

The pattern value itself (as opposed to its magnitude) can be a negative or complex number, 
encoding details about {\it how} the pattern is matched,
so it can ``interfere'' with other pattern values
when two boolean functions are composed 
and their pattern values are multiplied, 
as described in the next section.

\subsection{... which are entries in a pattern matrix ...\label{S:EntriesDefSubadd}}

The pattern values are the entries of a very large \emphdef{pattern matrix},
defined so that composing functions corresponds to multiplying these matrices;
this lets us prove our measured complexity formula
is a lower bound on the true circuit complexity,
by proving it's a ``subadditive matrix measure'' (at least for matrices of this special form),
and that it's correct for the circuit model's generating circuits
(e.g. all circuits containing at most one nontrivial gate, alongside any number of parallel wires).

By a \boldquote{subadditive matrix measure} in this context,
we mean any $C$ which takes pattern matrices
to values in $\R^{\ge 0}$ and for which $C(AB) \le C(A) + C(B)$.
If such a $C$ is applied to a circuit or circuit function, 
we treat it as being applied to the associated pattern matrix;
in that case we'd have $C(f \compose g) \le C(f) + C(g)$.
Calling this ``subadditive'' is an abuse of terminology,
justified by thinking of composition of circuits as being like addition,
at least when circuit cost or complexity is being discussed.\footnote{
In this paper we never need the conventional meaning of ``subadditive'' 
for a measure on matrices or functions.
In some other context where we did,
we might distinguish these terms by referring to this paper's version
as ``p-subadditive'' or ``product-subadditive''.
But for measures on the field $K$,
we {\it do} use ``subadditive'' with its conventional meaning.
}

One possibility for the choice of subadditive matrix measure (which we'll assume is our actual choice, in most of what follows) 
is simply the log of the ``sum of pattern magnitudes'',
with the ``sum'' itself actually being a ``sub{\it multiplicative} matrix measure''
with value at least 1;
such a measure might be as simple as the maximum number of nonzero entries in any row
(and for simplicity we'll assume it's exactly that, for most of our discussion).
(By a \emphdef{submultiplicative matrix measure} we mean any $M$ which takes pattern matrices
to values in $\R^{\ge 0}$ and for which $M(AB) \le M(A) M(B)$. 
This is a standard use of the term ``submultiplicative'' 
(except that we only require it to work for pattern matrices, not necessarily for all matrices).
For more examples and related theorems, see Section~\ref{S:Definitions}.)

(To clarify: the phrase ``sum of pattern magnitudes'' is meant to be suggestive, 
with ``submultiplicative matrix measure applied to matrix of pattern values'' being its precise version --- 
even though such a matrix measure would not just be a ``sum of all the magnitudes''.
It might be the maximum over rows of the sum of magnitudes in each row, 
or something more complicated; but like an actual sum (of nonnegative values), 
it could be lowerbounded by lowerbounding subsets of the terms.)

\subsection{... of very high dimension ...\label{S:ofveryhighdimension}}

The pattern matrix dimensions for an arbitrary boolean function $f$
from $n$ bits to $m$ bits 
are not $m$ by $n$,
or even $2^m$ by $2^n$ (like the matrices used to describe quantum computation, when $m = n$),
but $2^{2^m}$ by $2^{2^n}$,
since the matrix represents a linear map 
between arbitrary $K$-valued \emphdef{properties} (functions) of {\it truth tables} (not just properties of bit vectors).\footnote{
    This is the simplest of several possibilities, 
    but the others would also involve large dimensions 
    (see Section~\ref{S:DefPatternBasis}).
}

Specifically (as
demonstrated in Section~\ref{S:patmatforAND}, and
described formally in Section~\ref{S:DefPatternMatrix}), 
we reinterpret $f$ as mapping truth tables (presented to it on its output, describing arbitrary boolean functions from $m$ bits to 1 bit)
into other truth tables (visible on its input to whatever uses it, thus describing boolean functions from $n$ bits to 1 bit, 
created by composition of (the function computed by) the $m$-bit truth table with $f$).
We then ``linearize'' this
(so it maps formal linear combinations of truth tables (with coefficients in $K$) from $f$'s outputs to its inputs),
and finally dualize it to get $f$'s \emphdef{pattern map}, a linear map from $f$'s inputs' \emphdef{property space}
(vector space of arbitrary $K$-valued functions of truth tables with 1 output and $n$ inputs)
to its outputs' property space (the same, but over truth tables with $m$ inputs). These vector spaces (over the field $K$)
have dimensions $2^{2^n}$ and $2^{2^m}$ respectively,
so we represent the pattern map by a $2^{2^m}$ by $2^{2^n}$ pattern matrix (with entries in $K$), as said above ---
but using a specially chosen \emphdef{pattern basis} in each property space, discussed below, 
to make the matrix entries appropriate for the complexity formula.

(Note that we can't assume the pattern basis will be orthogonal,
relative to the standard \emphdef{indicator basis} 
(whose elements are properties valued at 1 for one truth table and 0 for all others).
If we refer to a specific correspondence between a pattern and its \boldquote{dual pattern}
(a formal linear combination of truth tables),
we mean the corresponding element in the dual basis of the pattern basis.)

\subsection{... so the log of a submultiplicative matrix measure can range up to $2^n$. }

A submultiplicative matrix measure (like the one which counts nonzero entries in rows)
can be made to have values that range from 1 to (approximately) one of the matrix dimensions,
so the large size of the pattern matrix allows the log of such a measure
to be roughly as high as the maximum possible circuit complexity of a function,
which (for circuit models which permit general boolean functions, 
and have not too many generating circuits, each with small cost)
is comparable to the number of bits in a truth table, $m 2^n$.
(The fact that a complexity measured this way could not get {\it quite} that high, 
especially since the pattern matrices should be sparse (as we'll discuss below),
is significant, but not a problem for some important applications; later we'll discuss what that fact might mean,
as well as giving more principled reasons for this formal setup than the size range of values it can produce.)

\subsection{This does prove lower bounds, but they're often trivial...\label{S:oftentrivial}}

If we choose any field $K$, and define a pattern basis as {\it any} fixed
choice of basis for each size of property space,
then for all choices of pattern basis
and many choices of submultiplicative matrix measure,
we could construct a ``subadditive'' function of $f$'s truth table in the sense described above,
and prove that it lower-bounded any $f$'s circuit complexity 
(after scaling it to fit the costs of the generating circuits of whatever 
numbers of inputs and outputs $(n, m)$ might be needed).

But for most (i.e. for ``random'') choices of pattern basis, this result would be useless, for two reasons:
\begin{itemize}
\item 
the \emphdef{arbitrary width issue:}
We don't know in advance how ``wide'' a circuit might be needed to compute $f$ 
(where for any circuit expressed as a composition of generating circuits $g_i$,
called its \emphdef{circuit stages},
by \emphdef{circuit width} 
we mean the maximum $n_{g_i}$ or $m_{g_i}$ 
of any circuit stage $g_i \colon \F_2^{n_{g_i}} \to \F_2^{m_{g_i}}$); 
but the ratio of (unscaled) measured complexity to actual circuit cost might grow arbitrarily high
as a generating circuit is ``widened'' by adjoining non-interacting wires, as our pure-serial-composition model requires us to do.
So to scale the measured complexity correctly for all possible widths, the scaling factor would often have to be 0.
(Technically, we could limit the width to the maximum possible circuit complexity for the whole function $f$, 
but in practice this would be just as bad.)
\item
the \emphdef{arbitrary basis issue:}
Even for a fixed width $w$, if the pattern basis is arbitrary relative to the generating circuits,
it will probably treat their functions like any other functions of the same \emphdef{size} $(n,m) = (w,w)$;
thus the resulting complexity measure will probably consider the generating circuits
as having almost maximal complexity for their size (as it does for most functions),
rather than as having especially low complexity (as we want it to do).
\end{itemize}

These reasons turn out to be related ---
addressing the ``arbitrary basis issue'',
well enough to make a useful pattern basis for {\it fixed} widths,
will also solve the ``arbitrary width issue''.
(To get ahead of the story,
the basic reason is that they must both be addressed by making sure that 
trivial width-increasing operations like ``adjoining a non-interacting parallel wire''
don't increase a circuit's measured complexity, at least not by too much.
We'll cover this in detail later on.)

But for now, the situation will be clearer if we focus
on the special case where all generating circuits are constrained to have the same width $w$,
so $n = m = w$ for all functions we're considering ($f$ and each $g_i$).
(This is natural for reversible circuits, but doesn't imply reversibility.)

\subsection{(... even for circuits of fixed width ...)\label{S:evenfixedwidth}}

In the context of fixed-width circuits ($n = m = w$ for all stages, for some constant $w$),
all we're saying about an arbitrary ``random'' pattern basis is that
there's nothing special about a generating circuit's function
compared to any other non-identity function,
and most functions will have near-maximal measured complexity relative to that basis
(since their pattern matrices won't happen to be sparse),
so the generating circuits probably will too.
(Since the measured complexity is easily proven ``subadditive'', 
we could still scale it to fit the generating circuits' costs
(unless some have 0 cost --- we'll discuss those later),
but this would just result in a trivial lower bound
which proves that computing $f$ requires at least one generating circuit.)

So does {\it any} basis (of the property space for width $w$) 
avoid seeing the generating circuits as highly complex?

For reversible functions,
at least two very different bases do avoid it,
but as pattern bases they're ``degenerate'' --- so symmetrical that they give {\it every} reversible function
a measured complexity of~0.
These are the \emphdef{indicator basis} (in which each element recognizes a single truth table),
and the \emphdef{monomial basis} 
(each element evaluates the XOR of a fixed but arbitrary subset of truth table entries,
representing the result in $\{1, -1\}$; these bases are discussed further in later sections).
Though these are as far apart as any two bases can be
(according to the standard inner product in the property space),
both of them are just {\it permuted} by every reversible function,
which is why they measure its complexity as 0.
(Even irreversible functions have simple actions on these bases,
but in different ways ---
the indicator basis measures {\it every} function's ``complexity'' as 0,
but the monomial basis measures positive ``complexity'' 
in irreversible functions which ``forget'' information about their input
(by mapping multiple inputs in $\F_2^n$ to the same output in $\F_2^m$).
Note that this means there is a ``subadditive'' measure related to ``forgetting'',
but doesn't imply it has anything to do with actual circuit complexity.\footnote{
    This measure is shown in Table~2 of Section~\ref{S:comparingchoices}
    as $C_{M,\P}$ for $M = M_{\nz}$ and $\P = \P_{\mon}$.
    What it actually measures is how much smaller a set of possible inputs (a subset of $\F_2^n$) can become,
    after $f$ maps it into some subset of $\F_2^m$.
    (This is of course maximized by the entire input set, 
    but expressing it as a maximum over subsets
    makes the nature of its subadditivity clearer.)
    Incidentally, this is an example of a natural ``subadditive'' measure 
    definable by this framework, 
    whose value can range up to almost $2^n$.
})

So it's easy to find a basis which sees most functions as high complexity (including the generating circuits),
or  one which sees all reversible functions as 0 complexity,
or even one which sees {\it all} functions as 0 complexity.
But if we want a {\it useful} basis, we'll need something more sophisticated.

\subsection{... at least when the pattern basis ignores the generating circuits.}

None of the example bases considered so far
depended at all on the generating circuits~--- 
so it's not surprising their treatment of those circuits was trivial or generic.
Before addressing how a pattern basis {\it should} take them into account,
it's worth pointing out one reason we know it will have to ---
the set of generating circuits may seem special to {\it us},
but to the property space they're more or less ``just another set of non-identity functions''.

To be more specific: 
there are many automorphisms on the set of all boolean functions from $\F_2^n$ to $\F_2^m$,
corresponding to permutations of their $2^n$ input values and of their $2^m$ output values.
These automorphisms map the functions computed by the generating circuits to almost-arbitrary 
other functions.
This means that even if we happen upon a ``good'' pattern basis in some intrinsic sense, 
it might be measuring a different kind of complexity than we're interested in.
For example:
\begin{itemize}
\item
It might be measuring
complexity relative to some arbitrary encryption scheme for the tuples of bits 
passing from each circuit stage to the next. 
Such a formula would claim correctly that most but not all functions have near-maximal complexity, 
but would disagree with us about which functions to consider simple.
But it would do this consistently, in the sense of being ``subadditive'' over function composition.
The functions it considered as having minimal complexity, 
though appearing to us as individually almost random,
would obey the same relations among themselves as
a more ``conventional'' set of generating circuits would ---
for example, typically
each of them would commute
with many of the others.
\item
Or it might be measuring complexity for a different primitive gate set, 
or using different gate costs,
or even for computation of a different nature. (Arbitrary maps from output to input truth tables
(not to mention linear maps between formal $\C$-linear combinations of them)
can represent some kinds of computation that aren't physically possible 
(and perhaps some that are physically possible, but not ``classical''),
which nonetheless could conceivably have a good measured complexity function
which this setup could represent.)
\end{itemize}

So far we've seen ways to choose the pattern basis to get
complexity lower bounds that are technically correct (after scaling), but trivial,
and only in extreme ways~---
most functions high complexity, 
or all {\it reversible} functions 0 complexity, 
or {\it all} functions 0 complexity.
What we need is something in between those extremes.
But it needs to be ``in between'' in just the right way~---
not only measuring complexities with a high ``dynamic range'',
but with the functions computed by the {\it generating circuits} 
being the ones it deems simplest~---
and for widening those functions
(by adding new non-interacting inputs, and corresponding outputs, to represent adjoined parallel wires)
not to increase their perceived complexity.

It's time to examine more carefully exactly what that would require, 
in terms of specific features of the pattern basis
which we might be able to understand how to achieve.

\subsection{We should choose it so generators' pattern matrices are sparsest ...\label{S:gensparse}}

Technically, by a \emphdef{pattern} we just mean ``an element of the chosen pattern basis'' 
(in any property space we're using).
But to make the pattern basis {\it useful}, we have to choose it correctly,
which means taking account of the generating circuits,
so the measured complexity can have low (but mostly nonzero) values for their functions, 
but high values for most functions.

This means
the patterns in the basis should not just be {\it any} properties,
but properties with special significance (when they match the truth table of some 1-output function $f$)
for predicting the values of other patterns in the composition of $f$ with
a generating circuit. Or to be more precise,
we want pattern matrices (when expressed in the chosen pattern basis) 
to be as sparse as possible~---
even for arbitrary functions (so most of their pattern values are 0, for the reasons given earlier),
but especially for simple (low-complexity) functions (so they have smaller matrix measures).
For the functions we want to consider simplest of all, but nontrivial~---
i.e., the functions computed directly by generating circuits~---
we want pattern matrices with only a few nonzero entries per row
(or only a few ``large'' entries, if we're using a matrix measure 
which can effectively ignore large numbers of ``small'' entries)~---
but not just 1 nonzero entry on every row (for every generating circuit),
or they (and thus every constructible function) will end up with a measured complexity of 0.

Note that it's ok if {\it some} generating circuits
(which we want to consider trivial), 
e.g. those which only permute the order of wires,
have only 1 nonzero pattern value per row,
and thus a measured complexity of 0,
as long as those circuits alone can't be composed to make too many different functions~---
since their compositions will also have measured complexity of 0.
(When those functions are reversible,
we can achieve this by making sure the pattern basis is symmetric relative to those functions,
so they just permute it (and perhaps also multiply its elements by roots of unity in~$K$).
That symmetry makes sense anyway,
since (given that cost model) actual circuit complexity has the same symmetry.
We'll discuss irreversible 0-cost functions later.)

There is an interesting unconventional circuit-cost model (for general boolean computation)
which considers a larger set of functions to have 0 cost ---
namely, all $\F_2$-affine functions ---
which results in the maximum possible complexity
being roughly the square root of its usual value \cite{BPP00}.
This may have technical advantages in creating a useful pattern basis.\footnote{
    Treating all $\F_2$-affine functions (i.e. all XOR and NOT gates) as ``free''
suggests looking for a pattern basis 
which is symmetric not only for the $n!$ wire permutations and $2^n$ wire-negations
(in the input space $\F_2^n$, whose coordinates correspond to ``wires''),
but for arbitrary $\F_2$-linear changes of basis of the input space
(of which there are roughly $2^{n^2}$).
It also seems natural in conjunction with the
``fourier matrix model of boolean computation''
we'll discuss
in Sections~\ref{S:specfunc1} and~\ref{S:linalg},
in which any $f \colon \F_2^n \to \F_2^m$ 
is viewed as an $\R$-linear map from XORs of inputs
to XORs of outputs 
(but with each XOR-value $b \in \F_2$ represented by $(-1)^b \in \R$),
which is representable by
a $2^m$ by $2^n$ matrix of real-valued fourier coefficients of $f$; 
these matrices multiply when functions are composed.
}

\subsection{... which has not yet been achieved...}

So we know we want a pattern basis which makes all pattern maps have fairly sparse matrices, 
and simple ones sparsest,
with the generating circuits' matrices almost as sparse as the identity matrix, 
but not quite (except for the 0-cost generating circuits).
But we don't know any systematic way to construct a basis like that ---
if we did, we'd immediately get our useful complexity formula,
assuming we understood the basis well enough to prove, 
for at least some functions of interest, that they matched enough of its patterns.

It's worth noting that if we could prove a useful pattern basis {\it existed}, 
in the sense of all rows of generating circuits' pattern matrices being sparse, 
then even if we only ``understood'' its ``simple'' patterns 
(treating most of the rest as beyond our specific understanding),
we could potentially use it to prove significant lower bounds,
by finding enough matching of simple patterns to truth tables of interest.
We'd have to understand only enough about the non-simple patterns 
to prove the generating circuits didn't map any pattern's \emphdef{dual} 
(i.e. the corresponding formal linear combination of truth tables, in the dual pattern basis,
mapped from circuit output to circuit input)
to a sum of too many other patterns' duals.
(Such a sum corresponds to one row of the pattern matrix, 
so we're still talking about minimizing the number of nonzero values on each row.)

It's easy to keep this measurement low for single gates (such as S-boxes),
simply because they have only a constant number of inputs; 
the hard part is to avoid increasing it (the number of nonzeroes on any row) too much,
as those circuits are adjoined to more and more non-interacting parallel wires,
to construct generating circuits with more inputs and outputs.
(We sometimes call this the \emphdef{adjoined-wire issue}.)

(Most of what we've said so far should apply to most kinds of circuit models;
but starting here, we'll be talking mainly about models intended for general boolean computation.
Other models might be analogous in many ways, 
but the effect on the property space of ``adjoining a non-interacting wire'' 
(and even more, the considerations about how that relates to a useful pattern basis)
would often be different than they are in the general boolean case we discuss below.)

\subsection{... since naively ``adjoining a wire'' doubles measured complexity ...}\label{S:tensorwithitself}

The simplest way to try constructing a pattern basis is by induction over the number of wires $n$.
That is, given a basis $P_n$ for the property space $Q_n$ over $n$ wires (of dimension $2^{2^n}$),
we somehow construct a basis $P_{n+1}$
for the property space $Q_{n+1}$ over $n+1$ wires (whose dimension is $2^{2^{n+1}} = (2^{2^n})^2$).\footnote{
    It would be more precise to speak of $n$ inputs rather than $n$ wires,
    and (below) of the measured complexity of functions
    rather than of circuits~--- but ``adjoining a wire'' seems easier to visualize in the context of circuits.
    By the ``measured complexity of a circuit'',
    we just mean the measured complexity of the function it computes,
    whether or not the given circuit is minimal for that function.
}

The most straightforward version of that construction
can be described 
(at the level of a given circuit's pattern map, 
between its input and output property spaces)
as simply
\emphdef{tensoring the circuit's pattern matrix with itself},
to get the pattern matrix for the same circuit ``widened by one wire''~---
that is, with one new non-interacting wire adjoined in parallel.\footnote{
    If we define each $P_{n+1}$ as $P_n \tensor P_n$,
    where $P_n$ is the pattern basis set 
    for properties of $n$-input 1-output truth tables,
    this also makes pattern matrices
    tensor with themselves, when one non-interacting wire is adjoined.
    In symbols (where function
    $f'$ is $f$ with one ``wire'' (new input and output) adjoined), we have (in this case)
    $\PM_{\P}(f') = \PM_{\P}(f) \tensor \PM_{\P}(f)$.
    (See Section~\ref{S:changeofpatternbasis}
     for what we mean by ``tensoring basis sets'' like $P_n$,
     and Sections~\ref{S:Notation} and~\ref{S:Definitions} 
     for other notation and definitions used here.)
}
(If we do this at every inductive level,
we can end up with the indicator or monomial basis
in each size of property space,
depending on our choice of basis for
the base case of 0-input truth tables.)

The problem with this construction
is that as each wire is adjoined,
the number of nonzero terms in each pattern matrix row is squared,
which doubles the measured complexity.
(If we generalized this framework to let the new wire have $r$ possible values instead of 2, 
this construction would produce the $r$th tensor power of the pattern matrix, 
and thus multiply its measured complexity by $r$
as each new wire was adjoined.)

If the new input wire had {\it not} been non-interacting, but 
(in addition to controlling the new output wire)
also affected the rest of the circuit~---
effectively making it imitate
one of 2 (or $r$) unrelated circuits of the same complexity 
(with the new input value choosing which one)~---
then this effect of multiplying the circuit's measured complexity by 2 (or $r$) would be exactly right.
But it's a fatal flaw when the new wire is supposed to be non-interacting.
In effect, this naive construction is treating every adjoined wire as if it affected everything in the circuit,
resulting in a new circuit which is twice as complex 
(since its truth table is twice as large, and ``presumed arbitrary'')~---
even when the new wire is actually non-interacting.

To fix this problem, the inductive construction 
has to be done in a more sophisticated way,
which recombines or ``mixes''
(by an appropriate change of basis)
the ``naively''-tensored smaller-basis elements,\footnote{
    I.e., $P_{n+1} = W_{n+1} \big( P_n \tensor P_n \big)$
    (where $W_{n+1}$ is 
    a square matrix we design, for each $n$). 
    If $f'$ is $f$ with one wire adjoined, their pattern matrices obey
    $\PM_{\P}(f') = W_{m+1} \, \big( \PM_{\P}(f) \tensor \PM_{\P}(f) \big) \, W_{n+1}^{-\!1}$.
}
in a way which, when the pattern matrices being tensored are the same,
doesn't increase (too much, or preferably at all)
the number of pattern matrix terms (per matrix row) 
needed to express any given generating circuit.

Finding a single change of basis (for each value of $n$ in this inductive step), 
which would fix this problem for every generating circuit at once,
would be hopeless, if it needed to work for arbitrary matrices.
But the pattern matrices and pattern basis are far from arbitrary,
and it seems plausible they can be given special properties which make this possible.
There are at least two points we can take advantage of:
\begin{itemize}
\item
As we'll discuss near the end of Section~\ref{S:kindsofpatterns} (about kinds of patterns), 
the pattern basis might be able to have high symmetry,
even though most of its elements are ``complex''~---
since {\it compared to arbitrary properties}, they can have {\it very low} complexity.
\item
Each 1-output function (truth table) of $f$'s input or output space,
when it's expressed (uniquely) as a sum of dual patterns,
can {\it also} potentially have high symmetry 
(e.g. with respect to permuting its coefficients in certain ways, 
along with multiplying them by roots of unity in $K$),
since there are only as many functions as dual patterns (i.e. as dimensions of the property space),
so by no means do arbitrary combinations of coefficients need to be allowed.
(Even considering the larger number of functions {\it between} the input and output spaces,
their number is more like the number of dual patterns than like its exponential,
so this point remains largely true.) 
\end{itemize}

\subsection{... but there's hope, since that's the {\it only} hard issue we need to solve.}\label{S:hopesince}

In a conventional circuit model for general boolean computation,
it turns out we only need the pattern basis to satisfy three criteria,
to be able to prove nontrivial complexity bounds for most functions
(though this won't guarantee we can do that for specific functions of interest):
\begin{enumerate}
\item
it's symmetrical for permutations of wires, so it sees them as having 0 complexity
(or perhaps, ``almost symmetrical'', so it measures their complexity as ``small''~---
note that the term ``complexity'' as used here just
means ``whatever value our formula measures'',
whether or not that relates to actual circuit complexity);
\item
it measures most functions as having significant complexity (like a random basis would);
\item
it handles the adjoined-wire issue 
(so adjoining wires leaves measured complexity unchanged, 
or at least doesn't change it ``too much'', as detailed below).
\end{enumerate}

A conventional circuit model's ``free'' operations include
not only permuting wires, 
but discarding or splitting them, 
or creating new (constant-valued) wires.
Ideally our pattern basis will agree 
(by giving those \emphdef{trivial wire operations} a measured complexity of 0);
but if not (as long as it meets the listed conditions),
we can work around this by artificially treating them (in a modified circuit model) as having a small constant cost,
which gives only a small linear weakening of our lower bounds.\footnote{
Technically this weakening is not precisely ``linear'', since a function with arbitrary $n$ and $m$
which can be computed using {\it only} trivial wire operations (e.g. discard all inputs, then output all zero bits)
would have 0 complexity in a conventional model, but arbitrarily high complexity
if trivial wire operations were given constant positive cost.

But if we ignore or work around this issue
(for example by first redefining conventional complexity to be higher by $n + m$),
then ``linear weakening'' is accurate:
a lower bound of $X$ on our revised complexity (with trivial wire operations of constant positive cost)
implies a lower bound of $\Theta(X)$ 
on the originally-defined (but with extra $n+m$) circuit complexity (and the suppressed constant factors are small).
Proof: a minimal circuit in the original cost, using $X$ gates, 
needs at most $n + m + cX$ trivial wire operations,
for some small $c$ depending only on the circuit model 
(namely, the maximum of $n_g+m_g$ over the primitive gates, where gate $g$ has $n_g$ inputs and $m_g$ outputs). 
The scaling factor change is also constant and small. The result follows.
(A more formal but much longer version of exactly the same proof can be found in
Section~\ref{S:fallbackproofs1}.)
}
However, I would guess that making those operations have 0 measured complexity in the first place
will be easier than solving the adjoined-wire issue, and may help show how to solve it
(since both issues involve the relationship between the pattern basis
and the subalgebras of the property space 
consisting of properties definable on various subsets of wires).

The reason we consider ``adjoined wires'' the main issue
is that solving the other two together doesn't look hard.
For each input or output space of size $n$ (namely, $\F_2^n$, whose $n$ coordinates correspond to ``wires''),
the wires have only $n!$ permutations,
so constraining an otherwise-arbitrary basis of the associated property space (of dimension $2^{2^n}$)
to be symmetrical under their action should still leave it a lot of freedom, for example to be ``mostly random''.

(Of course, solving the adjoined-wire issue would presumably force the pattern basis to be highly {\it non}random;
but the fact that {\it most} bases would give most functions high measured complexity
gives reason to believe that
that feature (number 2 in the list above) is not impossibly constraining either.)

\subsection{Proof outline: a pattern basis meeting {\it these} conditions will be useful.}\label{S:proofoutlineCMPlambdaexists}

Here is a more precise version 
of the statement about the ``three criteria'' made above, 
and a preview of the proof
(which is given more formally in Section~\ref{S:idealcase}):

\bigskip
Suppose we construct a pattern basis which fixes the adjoined-wire issue in the ideal way ---
so adjoined (parallel, non-interacting) wires don't affect measured complexity at all, at least for generating circuits ---
and for which permuting wires has 0 measured complexity (due to symmetry).

In a conventional circuit model for general boolean computation,
we can get by with just one kind of generating circuit (aside from wire permutations):
one small circuit (which might be either a primitive gate or a trivial wire operation)
with zero or more adjoined wires.

The small circuits have low measured complexity
(since their maximum size is a small constant of the circuit model),
so by our assumption about the effect of adjoined wires, the generating circuits do too ---
their measured complexity
is limited by a small constant.
(This also solves the ``arbitrary width issue''
(about scaling measured complexity to fit the costs of the infinite set of generating circuits of arbitrary widths),
as promised a few subsections back, 
since it means we only have to take into account the finite set of small circuits that form their ``cores''.)

Since all generating circuits have small measured complexity,
a small positive constant $\lambda$ (dependent only on the circuit model and pattern basis)
will scale that to fit their actual cost --- 
and after scaling, measured complexity remains ``subadditive'' for function composition,
so it becomes a valid lower bound on circuit complexity for all constructible functions in the model.
Then for {\it every} function $f$ with high measured complexity in this pattern basis,
we immediately get a proportionally-high lower bound on $f$'s actual circuit complexity.

(So, informally, a pattern basis meeting the above conditions
will be ``useful'' if we can prove it measures high enough complexity
on enough functions $f$, or enough ``interesting'' functions $f$, to consider that an interesting result.
The above conditions guarantee its measured complexity
for {\it all} constructible functions $f$ 
is (up to a small constant factor) 
a lower bound on $f$'s actual circuit complexity.)
\endproof 

\bigskip
(Even if we can't solve the adjoined-wire issue in the ideal way ---
that is, if adjoining a wire adds {\it some} measured complexity ---
we're ok for many purposes
as long as the amount added can just be a fixed polynomial in $n$,
so adding it needn't double the measured complexity of the original circuit.
Similar comments might apply even if permuting wires is not free.
We'll further discuss these ``fallback'' situations in Section~\ref{S:fallbackproofs}.)

\bigskip
\bigskip
\noindent As a possible guide for finding a useful pattern basis, I'll describe in more depth
the nature of the patterns we hope to get
by choosing the pattern basis ``correctly'',
and how they might relate to the kinds of patterns we understand intuitively.
(In later sections I'll describe 
the considerations which motivated this framework,
and give more reasons we might expect finding a useful pattern basis to be possible.)

\subsection{desired qualitative nature of patterns}

This subsection summarizes the general picture I hope can be found,
of the kinds of patterns that might exist in a useful pattern basis (for general boolean computation),
how various kinds of functions would match them,
and
how enough provable matching
might be found
to lower-bound the complexity
of interesting functions.
(Later sections go into more detail about some of these points.)

\subsubsection{kinds of functions\label{S:kindsoffunctions}}

\begin{itemize}

\item
In general, a function $f$ of measured complexity $X$ 
matches about $c^X$ patterns for some smallish $c$ 
(which I guess is constant, but we could survive if it was linear or even polynomial in $n$). 
By $f$ \emphdef{matches} (or \emphdef{has}) pattern $p$, 
we mean the associated pattern value is either nonzero or non-``small'', 
depending on the matrix measure used to construct the complexity measure. 
(Nonzero seems more likely than non-``small'', 
due to the quantized nature of the truth table and therefore of
the possible pattern values (especially if patterns have high symmetry as properties, as seems likely);
and if $K$ is a finite field rather than $\R$ or $\C$,
it's not obvious what ``small'' could mean 
(though something involving subfield membership might be possible).
I think this means ``nonzero'' should be hoped for here,
but it doesn't seem safe to completely assume.)

\item
The patterns differ in their complexity 
(for a meaning of ``complexity'' which is not fully precise here,
but for which Kolmogorov complexity of their expression
as a polynomial in the truth table entries 
(noting that the entry indices have structure which matters here) can stand in for now, 
since we essentially only use it to avoid having to make explicit counting arguments);
we say more about the nature of individual patterns below.
An important point mentioned earlier 
is that any algorithm to probe an arbitrary truth table at a polynomial number of possible patterns
(relative to the size of the truth table, about $2^n$)
is unable to scan non-simple patterns with any significant ``density'' (the fraction which get scanned).

\item
Low-complexity functions with special structure might have a predictable set of simple patterns. 
In some cases this lets us prove a lower bound on their complexity, 
by identifying enough patterns we can prove they have.

\item
High-complexity functions with special structure might or might not have a predictable set of complex patterns 
(as well as some simple ones). 
Again, if we can prove they have enough patterns,
we can prove a nontrivial bound on their complexity.

\item
Pseudorandom functions are low complexity and thus have relatively few patterns,
but (essentially by definition of ``pseudorandom'')
those patterns are all somewhat complex (though not maximally complex),
and are predictable only if you know exactly how the functions are constructed
(which, again by definition of ``pseudorandom'' \cite{GGM86},
must include complete knowledge of their construction parameters,
including ``random seed'', not just of the general scheme).
Thus it's hard to notice their relative lack of complexity 
(or anything specific about what kind of complexity they have)
by probing either simple or randomly chosen patterns.
(You could also just ``look at the truth table and see if you notice anything'',
but, we believe, this effectively probes only some simple patterns, not any complex ones.)
(Presumably there is some theorem that lets you infer,
solely from some function $f$'s complete lack of simple patterns,
that it must have at least some minimal complexity (probably roughly linear in $n$), 
even though you can't find any of the many patterns that theorem proves it must have.)

\item
``Random'' functions (those of near-maximal circuit complexity, as is true of most functions)
have {\it more} patterns than any other kind of function (and almost all of their patterns are very complex),
but still there are far more patterns they {\it don't} have
(i.e., as with all functions, most of their pattern match values are zero or ``small'').

\end{itemize}

(A few patterns might be not only ``simple'' but \emphdef{non-discriminating},
i.e. they might match a substantial fraction of all functions,
so they would be an exception to the general but approximate statement that a random or
pseudorandom function matches {\it no} simple patterns.)

\subsubsection{kinds of patterns\label{S:kindsofpatterns}}

\begin{itemize}

\item
Some of the patterns in a useful pattern basis would be simple ones we might recognize.
They might include (or in some sense ``correlate to'')
the kinds of facts about a truth table for $f$
which often have ``visible consequences'' when $f$ is composed with some generating circuit ---
that is, which often lead to other patterns being visible in the truth table of the composition ---
such as $f$'s \boldquote{density}, or its correlation with specific very simple truth tables,
or (more generally) the values of low-degree highly-structured polynomials in its entries.

For this purpose it's often convenient to treat truth table entries 
as elements of $\{1, -1\}$, i.e. as $(-1)^b$ for $b \in \F_2$;
we then have another basis of truth table properties consisting of monomials in those entries,
each of which is the XOR (product, in this representation, corresponding to the sum in $\F_2$)
of some fixed subset of entries. 
Using this representation,
the density of a truth table is just the average value of all entries, 
and its correlation with any fixed truth table is a degree-1 polynomial in the entries; 
the $2^d$th power of the $d$th Gowers uniformity norm\footnote{
That is, $\|\cdot\|_{U^d(\F_2^n)}^{2^d}$;
this was discussed as a property of boolean functions in \cite{Gow09}.
The norm $\|\cdot\|_{U^d}$ 
was introduced in \cite{Gow01}; 
see \cite{WikG} for notation.
},
and the ``influence'' of any set of $d$ inputs \cite{KKL88},
are highly-structured degree-$2^d$ polynomials.
(A caveat: these examples, and some others in this subsection, 
assume the property space is based on a field $K$ 
in which density and correlation values can be expressed directly, 
i.e. $\R$ or $\C$.
If $K$ was a finite field (and especially if its characteristic was 2), other examples would be more suitable,
though some of them taken in combination 
could still correlate to the kind of property normally expressed with values in $\R$.)

\item
An interesting class of simple patterns are the $2^{m+n}$ \emphdef{fourier coefficients}
\cite{Odo14},
each defined as the expected product (under the $(-1)^b$-representation)
of $f$'s output (or the XOR of some subset of its outputs)
and the XOR of some subset of its inputs.
These can be organized as a \emphdef{fourier matrix} 
which represents $f$ as a linear map from $\R^{2^n}$ to $\R^{2^m}$
(which we'll discuss later as the ``fourier matrix model'').
This involves only tiny subspaces of the corresponding property spaces 
(of dimension $2^n$ rather than $2^{2^n}$),
but it's worth pointing out that within them,
we can already do 
a ``miniature version'' of what we want to do in the property space as a whole.
Specifically, the fourier matrix of a $w$-input gate (plus any number of adjoined wires)
has at most $2^w$ nonzero entries per row (out of $2^n$);
as we compose circuits that number multiplies, so it grows at most exponentially
(a similar point was made in \cite{Gow09a}).
This lets us prove (trivial) circuit size lower bounds of the form $\log (2^q / 2^w) = \Omega(q)$, 
where $q$ depends on $f$ but can be no more than $n$.
If we could extend this behavior to the whole property space, those bounds could instead be more like
$\log (2^{2^q} / 2^{2^w}) = \Omega(2^q)$, where $2^q$ can be no more than $2^n$.
So in a sense, all we need to do is choose all the higher-degree patterns
so they share this important feature of the (degree-1) fourier coefficients.\footnote{
    This doesn't mean the fourier coefficients themselves must be elements of a useful pattern basis,
    but it seems likely 
    some patterns would be closely related to them.
}

\item
Other properties can be made by combining those patterns (using sum and product, i.e. forming polynomials of them),
which (if those polynomials are either small, or highly structured)
might exist directly in the pattern basis, or might be relatively small linear combinations of patterns in it.
Properties like this 
can express things like ``$f$ approximates a certain function of a few of its inputs'' 
(or ``of a few XORs of subsets of its inputs'')
---
the kind of pattern whose presence in $f$'s truth table intuitively implies the need for any circuit computing $f$
to somehow do ``a certain piece of computational work''.

\item
It's even possible to approximate {\it any function} of simpler patterns (not just a polynomial of them)
using relatively few terms in linear combination, by means of a short Taylor series. 
It's also potentially possible for any expressible property to be included in the pattern basis directly ---
for example, a property like ``log of the fraction of all truth tables which have the same density as this one'' ---
provided not too many are included, and they're linearly independent in the property space.
If there is some reason
for properties like that to be in the pattern basis, or to be expressible as small linear combinations of its elements,
we might be able to choose it to make that true. 
(For reasons like these, the pattern basis seems likely to be either closed under product of patterns
(where that product is defined as the pointwise product of pattern values, treating patterns as functions of truth tables ---
note that this is not the same as {\it composition} of patterns (when the functions containing them are composed),
though in other contexts we might also treat that as a product), 
or ``approximately closed'' in the sense that products have relatively small expressions as linear combinations
of pattern basis elements.)

\item
But most patterns (not only in the pattern basis, but in {\it any} basis of a property space) would be very complex;
note that there are $2^{2^n}$ patterns, 
so most of them have a Kolmogorov complexity of nearly $2^n$ bits --- as much as a random truth table.

An important observation related to this:
Any pattern $p$ which happens to match some random function $f$ (where $f$ is random in the sense
 of having near-maximal Kolmogorov complexity), 
but which is non-matching on almost all functions,
{\it needs} a high Kolmogorov complexity itself, or the very fact of its matching
will force $f$ to be non-random, since otherwise we could ``compress'' $f$ by describing it as 
``the $k$th function which matches $p$''. Or more precisely, when we {\it do} describe $f$ that way, 
it must be true that $C_K(k) + C_K(p) + C_{K0} \ge C_K(f)$, 
where $C_K()$ denotes Kolmogorov complexity and $C_{K0}$ is a small constant;
but $C_K(k)$ can't be too large a fraction of $C_K(f)$ if $p$ is non-matching for almost all $f$
(since then $k$ can't be nearly as large as the number of possible functions $f$).
Thus the only way to realize the vision of random functions being matched by {\it any} highly discriminating pattern,
not to mention by {\it lots} of such patterns, is to have lots of high-complexity patterns.

\item
On the other hand, relative to an {\it arbitrary} property of an $n$-bit truth table
(of which there are $k^{2^{2^n}}$ whose values are in any given $k$-element subset of $K$), 
the elements of the pattern basis can all be ``exceptionally simple''.

\end{itemize}

The last point
means we have plenty of ``room'' to be very selective about which properties we include
in the pattern basis --- for example, it may well be able to be a highly symmetrical structure
made only of special kinds of properties,
in spite of most of them being as complex as a random truth table.
Indeed, as mentioned earlier, we want it to include only those properties (of a function $f$) which 
are the most significant for predicting {\it other} patterns in compositions of $f$ with generating circuits~---
but which are ``significant in independent ways'', 
meaning not only that the patterns are linearly independent in the property space,
but that relatively few pairs of them interact directly,
in the sense of having nonzero entries in the pattern matrices of simple functions.

That criterion for a good pattern basis is both strong and self-referential
(though it's not vague, when expressed as pattern matrix sparseness for all simple boolean functions);
but the relative hugeness of the property space as a whole
would seem to give some hope that it can be met.

\subsubsection{lower-bounding the complexity of specific functions\label{S:specfunc1}} 

Assuming a good pattern basis could be found,
how might we prove a reasonable lower bound on measured complexity for a function with special structure?

To be more specific: 
given a well-understood function family $f = \langle f_n \rangle$, where
$f_n \colon \F_2^n \to \F_2^m$ has ``intuitive complexity'' $X$ (and $m$ and $X$ depend on $n$),
how might we prove $f_n$ has on the order of $c^X$ patterns 
(for some small constant $c$)?
(For this discussion, it's sufficient to assume $f$ has just one output (i.e. $m = 1$). 
It's also worth noting that even if we find only $c^{\sqrt X}$ or $c^{X^\alpha}$ patterns (for some $\alpha > 0$)~---
due either to limitations in
our pattern basis,
or to 
$f$ being less complex than we think~---
we could still get significant bounds like $\NP \nsubseteq \Ppoly$,
provided $X$ is superpolynomial in $n$ and $f$ is in $\NP$.)

\bigskip
As an example,
we'll let $f$ be a \boldquote{feature detector},
whose output is true whenever its input matches any of $X$
seemingly independent ``features'' $F_i$ 
(predicates over $f$'s input space $\F_2^n$).\footnote{
    As we'll see after defining $\langle F_i \rangle$, $f$'s {\it true} complexity
will turn out to be at most $\widetilde{O}(\sqrt{X})$.
(Later we'll mention
a possible partial explanation for this discrepancy.)
This needn't discourage us from trying to prove a superpolynomial lower bound,
since $\sqrt{X}$ is superpolynomial whenever $X$ is.
}
Since we want $X > n$,
not all combinations of input features can occur~--- 
they can't be {\it jointly} independent.
Even so, by making them
{\it pairwise} independent 
(and giving them other properties we'll define below),
we'll be able to make an intuitive argument for $f$ having high complexity.

Though we can't formalize this argument (let alone prove it),
it will suggest a way in which 
$f$ might be able to match
enough patterns in a useful pattern basis.
Specifically, we'll argue that the following steps are plausible:
for each input feature $F_i$, 
find a set of $c$ 
properties $P_i$,
each of which $f$ matches due to its ``approximately detecting'' feature $F_i$;
observe that $f$ matches all $c^X$ {\it products} of those properties
(since the product of their (nonzero) values at $f$ is nonzero),
and argue that many of those product properties are distinct;
finally (using our guesses about the nature of a useful pattern basis),
derive from these a sufficient number of related {\it patterns}, also matched by $f$.

(Note that we're not suggesting this is the {\it only} way
to prove an explicit function matches lots of patterns,
nor are we claiming it can surely be made to work.
We're just arguing that a conclusion like this might be plausible,
by speculating on one possible way of reaching it.)

\bigskip
When thinking about possible patterns,
or reasoning (intuitively) about complexity,
we'll consider \emphdef{input-XORs} (XORs of subsets of $f$'s inputs)
to be just as fundamental as single inputs.
In other words, we won't pick a preferred basis in $f$'s input space $\F_2^n$
(considered as a vector space over $\F_2$).
This will help us define a superpolynomial number of input features,
since 
(over
input vectors in $\F_2^n$ chosen uniformly at random)
all pairs of the $2^n$ input-XORs are statistically independent. 

Our main justification for this viewpoint is the 
\emphdef{fourier matrix model of boolean computation}
(which we'll
also discuss in Section~\ref{S:linalg};
this should not be confused with the exponentially-larger \boldquote{property space model},
which is a major subject of this paper\footnote{
  Though we won't use this ``fourier matrix model''
except to support this section's intuitions about feature-detector functions,
it's worth pointing out that it's {\it related} to the property space model,
in the sense of being a ``submodel'' or ``projection'', obtainable
by ignoring all properties except 
degree-1 (homogeneous) polynomials in truth table entries (represented as $(-1)^b$)
(provided we're using property values in $\R$ or $\C$).
This raises the possibility that the vague ways we'll use it here
might be more successfully formalizable 
if we could extend them into the larger model.
}).
In the fourier matrix model,
we represent all bits 
$b \in \F_2$
(whether they're inputs, outputs, or boolean function values) 
by $(-1)^b \in \R$,
so XOR is multiplication in $\R$,
and the correlation between two \emphdef{signals}
(arbitrary boolean functions over $\F_2^n$)
is their expected product (over random input vectors).
This lets us represent any 
boolean function or circuit 
(from $n$ inputs to $m$ outputs)
by a linear map (from $\R^{2^n}$ to $\R^{2^m}$)
between its input-XORs and output-XORs
(or equivalently, by a matrix of all its fourier coefficients;
composing functions composes the linear maps and multiplies their matrices).\footnote{
    \cite{Odo14} also discusses using this $(-1)^b$-representation
    to compute fourier coefficients, XOR, and correlation
    of 1-output boolean functions,
    and the fact that the $2^n$ parity functions (what we call input-XORs)
    form an orthonormal basis of (what we call) the ``vector space of signals over $n$ inputs'', $\R^{2^n}$.
    But I'm not aware of any prior publication of the fourier matrix model itself,
    or of any representation of boolean function composition by matrix multiplication,
    except for the special case of reversible functions,
    which can also be considered a special case of quantum computation \cite{NC00}
    (an example of this correspondence, involving fourier coefficients,
    was noted in \cite{Gow09a}).
}

This model lets us see any circuit $C$ which computes $f$
as some way of {\it linearly} moving and combining
$C$'s input-XORs,
placing them in turn onto the wire-XORs 
of $C$'s intermediate wires at each ``stage'' (single gate, and all wires alongside it),
so as to end up with just the desired combination of input-XORs 
on $C$'s output~---
namely, the fourier components of $f$.
To the extent that $f$ approximates some function $f'$
of the input-XORs in some subspace $W$ of $(\F_2^n)^*$ 
(the dual of $f$'s input space, as a vector space over $\F_2$~---
this is just the space of all the input-XORs of $f$, or of $C$),
we can argue (intuitively),
after noting that all fourier components of $f'$ are
(represented by) 
elements of $W$,
that the only ``computational work'' (linear motions/combinations of signals within the circuit)
which is ``useful'' for $f$'s approximation of $f'$
is the part involving
the input-XORs in $W$
(as potential fourier components of any wire-XOR in the circuit~--- that is,
if we were to formalize ``this part of the work'',
the first step would be to project every signal in the circuit
into the part containing only those fourier components).

Though we can't fully formalize ``computational work'' or our reasoning about it,
this viewpoint suggests intuitively
that if some circuit $C$ simultaneously
approximates two functions $f'$ and $f''$ (in a way we'll describe more precisely below),
each determined by the input-XORs in independent $\F_2$-subspaces $W'$ and  $W''$ of $(\F_2^n)^*$,
then there is a limit to how much the ``work'' needed for $C$ to approximate each of $f'$ and $f''$
can ``overlap'' (make use of the same ($\F_2$-subspaces of) wire-XORs in $C$). 
This is because $f'$ and $f''$ are statistically independent
(over random input vectors of $C$),
and so are any other functions of the input-XORs in $W'$ and  $W''$ respectively;
but those are the only intermediate functions we recognized as ``useful'' for their respective computations
(getting $C$ to approximate $f'$ and $f''$),
and which we therefore defined as part of that ``work''.
But any one wire-XOR's ability 
to be correlated with independent signals
(such as all nontrivial input-XORs in $W'$ or $W''$ respectively)
is limited, since the sum of the squares of its fourier coefficients is limited to 1.\footnote{
    The fact that this limit is on a sum of {\it squares}, rather than of {\it absolute values},
is one possible source of the discrepancy mentioned earlier 
between our example function's ``intuitive complexity'' of $X$
and the known upper bound of $\widetilde{O}(\sqrt{X})$ 
on its actual complexity.
}

The above informal reasoning is far from conclusive;
as far as I know, there is no easier way to usefully formalize it
than to somehow extend it into the exponentially-larger property-space model.
But it does suggest the following construction of our example function.
After spelling that out,
we'll slightly improve our justification of its ``intuitive complexity'',
and then discuss it in relation to a hypothetical pattern basis.

\bigskip
To construct specific feature predicates $F_i$
for our example function family $f = \langle f_n \rangle$,
we'll use parameters $w$, $k$, and $X$
to be chosen later (as functions of $f$'s input size $n$).
Abstractly, we'll follow these steps:
\begin{enumerate}

\item
Choose $X$ pairwise-independent $w$-dimensional subspaces 
of $(\F_2^n)^*$.
We'll call the subspaces $W'_i$; 
in each one,
choose a basis $W_i$, which will be
a $w$-element set of input-XORs of $f$.

\item
Define each $F_i$ as some fixed function of the (values of the) $W_i$, 
which is true (i.e., the $i$th input feature is present) for only one combination of values
(so for random input vectors, it will be true $1/{2^w}$ of the time).
Choose parameters so that for most input vectors, every $F_i$ is false (and thus $f$ is false).

\item
To ensure $f$ is in $\NP$, 
it's enough to do all this using
a systematic structure, 
so we can nondeterministically guess an ``index'' $i$,
and efficiently check whether the
$F_i$ it indexes is true.

\end{enumerate}
To do this concretely,
organize the $n$ inputs of
$f$ 
into a matrix with $w$ rows
(ignoring ``leftover'' inputs if $n$ is not a multiple of $w$);
then
define $f$ as true
whenever some subset of $k$ columns
(each treated as a ``column vector'' in $\F_2^w$)
XORs to the 0 vector
(where XOR means vector addition in $\F_2^w$).\footnote{
    This is reminiscent of the ``d-SUM problem''
    (in which the column vectors would instead be added as signed
     $w$-bit integers), which is well-known (according to \cite{PW10}).
}
Or in the terms we used above:
each index $i$ picks some subset of $k$ columns in the matrix of inputs;
the associated set $W_i$ contains, for each of the $w$ rows of that matrix,
the XOR of the $k$ inputs in
that row and the $k$ chosen columns;
each $F_i$ is true whenever all $w$ of those XORs 
(i.e., all elements of the associated set $W_i$) are 0.

The number of input features $F_i$ in this construction
is $X = \binom{\lfloor n/w \rfloor}{k}$. If $k$ slowly increases with $n$
and $n \gg w \gg \log_2 X$
(for example, $k = \log n$ and $w = k^3$, rounding up to integers as needed),
then $X$ is superpolynomial in $n$, but for most input vectors every $F_i$ is false.
(As alluded to earlier, there are circuits of size $\widetilde{O}(\sqrt{X})$
which can evaluate this function,
by sorting
the XORs of each subset of $k/2$ column vectors, 
then looking for repeated values (which will be adjacent in the sorted list) 
that don't involve overlapping column subsets.)

We can now make more precise (though it will remain informal)
the sense in which $f$ ``approximately detects'' each input feature $F_i$,
and therefore (intuitively) 
seems to require its circuits to 
do ``most of the computational work'' needed to fully detect that feature
(and furthermore, that this ``work'' is pairwise ``independent'' for $i \ne j$).

Imagine repeatedly running $f$ on random input vectors chosen uniformly from $\F_2^n$,
and asking how close it comes
to computing a given $F_i$.
Since $F_i$ depends only on
the values of the $w$ XORs in $W_i$, we'll call them
``important input signals'', and call
everything else about $f$'s input vector ``random noise''.
We might then ask:
``conditioned on the values of the XORs in $W_i$ 
(with the input to $f$ otherwise random), 
how good is $f$ as a measuring device for $F_i$?''
(Note that even if we 
extend our definition of ``important'' input signals, by ``XOR-closure'',
to cover
the entire $\F_2$-subspace
$W'_i$,
so that $F_i$'s ``$(-1)^b$-representation'' 
is an $\R$-linear combination of 
``important'' signals' $(-1)^b$-representations,
this won't affect what we're conditioning on,
since the extended signals are functions of the original ones.)

The answer is, $f$ is a {\it very} good measuring device for $F_i$~---
for {\it any} specified set of values of the $w$ XORs in $W_i$,
the probability that $f$ and $F_i$ disagree
is bounded above by the (conditional) probability
that any $F_j$ (for $j \ne i$) is true,
which (by the union bound)
is at most $\sum_j 2^{-w} = (X-1) / 2^w \ll 1$.
(Note that the pairwise independence
of
the $W'_i$
is crucial for this conclusion,
since it tells us that conditioning on values of $W_i$ has no effect on values of $W_j$.)

That completes our intuitive argument for any circuit $C$, which computes $f$, necessarily
``doing most of the work'' of detecting each $F_i$.
But the same argument also shows
statistical independence
between {\it any} function of the $W_i$ 
and {\it any} function of the $W_j$ (for any $i \ne j$),
which is just the intuitive argument we described earlier 
for the ``work'' associated with approximately detecting $F_i$ 
being  ``independent''
of the similar ``work'' associated with~$F_j$.

\bigskip
Now we turn to the question of finding patterns in our example function $f$.
For this, 
we'll leave aside the vague concept of ``computational work'',
but replace it with the hypothetical (though not as vague) assumption
that we've found a ``useful'' pattern basis
with some of the properties we hope for,
especially in connection with how it relates to
the vector space structure of
$f$'s input space $\F_2^n$,
and how it solves the ``adjoined-wire issue''.
In light of our discussion above,
we'll 
assume our pattern basis is symmetrical 
(up to scalar multiples)
with respect to any change of basis in $\F_2^n$ or $\F_2^m$,
as well as any negation of inputs or outputs~--- in other words,
with respect to composing $f$ with any $\F_2$-affine function.\footnote{
    See also our discussion of \cite{BPP00} and related topics,
    in Section~\ref{S:gensparse} and its footnote.
}

Near the beginning of this subsection,
we outlined steps we hope to show are plausible
for finding on the order of $c^X$ patterns in $f$.
The first step is to find, for each input feature $F_i$,
a small set of properties $P_i$,
which $f$ has
due to its approximate detection of $F_i$.
In light of our discussions above and in Section~\ref{S:kindsofpatterns},
it should be clear that for any property of $F_i$ itself 
(considered as a function of the smaller input space $\F_2^w$ corresponding to the values of the $W_i$)~---
or more precisely, for a property of functions from $\F_2^w$ to $\R$ 
(which we can apply to the expected value of $f$, conditioned on the values of the $W_i$)~---
there are many related properties of $f$
(which measure things related to how well, in various senses, 
$f$ approximates various functions of the $W_i$);
for example, various sums and products of 
(the properties which measure)
the $2^w$
fourier coefficients of $f$
which are nonzero in $F_i$.
These provide many candidates for elements of $P_i$.

Next, we observed that if $f$ has several properties $\langle p_i \rangle$,
it also has their
pointwise product\footnote{
    This ``pointwise'' product of properties is defined 
    by treating them as functions of truth tables; it should
    not be confused with the product (of dual properties) coming from function composition.
}
property $\prod_i p_i$.
(The reason is simply that all it means for $f$ to ``have'' a property $p$
is for $p$'s value at $f$ to be nonzero.)
This already gives us $c^X$ properties which $f$ has 
(where $c$ is the minimum size of any $P_i$).
(These product properties might not all be distinct, but to the extent
that the values of all $W_i$ taken together can distinguish many input vectors from one another,
it seems likely that enough of them are distinct.)

By itself, this is not yet interesting~--- for any arbitrary property,
$f$ is likely to have it.
What's interesting is finding lots of {\it patterns} $f$ has,
not just lots of properties. 
What we need is some connection between the properties we've mentioned,
and the hoped-for structure of a useful pattern basis.

Recalling our discussion of the adjoined-wire issue, 
we hoped to solve it,
as part of the inductive construction of a pattern basis,
by an appropriate change of basis in a property space
which recombined patterns defined in its subspaces in a useful way.
Combining this with our hoped-for symmetry of a pattern basis
with respect to changes of basis in $\F_2^n$,
something similar should hold for the relationship
between patterns in the property spaces for $\F_2^n$
and for the (dual) subspaces $W'_i$ 
defined above.
The hope is that, once a specific useful pattern basis is understood,
its relationship to candidate properties for these sets $P_i$, and their product properties,
will be simple enough to find enough patterns in $f$
related to sums of those product properties.

Of course this hope (not to mention the evidence for its being achievable) is quite vague;
but I think these arguments show that it's at least plausible
(that is, not obviously impossible)
that an explicit function family similar to this example
could provably have
enough patterns to give us a nontrivial lower bound on its complexity.




\newpage 

\subsection{Figures and examples\label{S:Figures}}

These subsections give examples of pattern matrices,
and illustrate how they can be constructed and used.
They can be skipped if desired,
since most of
the same information is described elsewhere,
both informally (earlier in this Summary), and formally
(Sections~\ref{S:Notation} and~\ref{S:Definitions},
which also define some notation we use here).

\subsubsection{computing measured complexity ($C_{M,\P}$)}

The table on the next page,
read from the top down, 
gives an overview 
of computing the pattern matrix, 
and from it the measured complexity,
for some tiny example functions,
each of the general form $f \colon \F_2^n \to \F_2^m$
(that is, a boolean function from $n$ inputs in $\F_2 = \{0,1\}$
to $m$ outputs in $\F_2$).

Each function is described in the table by 
a circuit diagram, and equivalently by
a tuple of output formulae; these are
boolean formulae in its inputs $x$ and $y$.
(For example, the two-output function ($\logicneg{x}$,~$y$) 
sets its first output to the logical negation of its first input $x$,
and sets it second output to a copy of its second input $y$.)

In this table, the functions all have exactly 2 inputs, and either 1 or 2 outputs.
If these functions were part of a larger circuit, 
they would have to be ``widened'' 
by adjoining non-interacting parallel wires
to account for all wires ``passing alongside them'' in that circuit;
this would give them more inputs and outputs,
and thus make their pattern matrices larger.

The function shown in the left column ($\logicneg{x} \logicand y$) is the composition
of the two functions to its right~--- or as the table shows in symbols,
$(\logicneg{x} \logicand y) = (x \logicand y) \compose (\logicneg{x}, y)$
(with the last function, $(\logicneg{x}, y)$, having two outputs).
This lets the table's lower rows 
show the relationships 
implied by that composition,
such as matrix multiplication equalities,
or the subadditivity of measured complexity which results from that.
(The table also shows ``fourier matrices'' (Section~\ref{S:specfunc1}) for comparison, 
though they're not used for computing measured complexity.)

We show the measured complexity $C_{M,\P}(f)$,
defined as $\log_2\,M(\PM_{\P}(f))$,
for one choice of pattern basis $\P$, 
and two
choices of matrix measure $M$.
As we've explained elsewhere,
its value depends on $M$ and $\P$,
and (though always subadditive)
is not necessarily related to actual circuit complexity.
To become a lower bound on actual complexity, 
it must be scaled to fit generating circuit costs (of all allowed circuit widths);
for this to be possible, and to 
result in nontrivial bounds,
requires a carefully chosen pattern basis
(whose existence has not yet been proven for $m$ or $n > 2$).

The pattern basis used here, $\P_{\RmTwob}$,
is only defined (so far) up to $m,n \le 2$.
(See Section~\ref{S:egpatbases} 
for more information about $\P_{\RmTwob}$,
and Section~\ref{S:comparingchoices}
for a comparison with other pattern bases, some of
which can be defined for any $m$ and $n$.)

\newpage

\begin{center}
\textbf{Table 1:} how measured complexity is computed (and why it's subadditive)
\end{center}

\renewcommand{\arraystretch}{1.5} 


{ \small


\noindent \begin{tabular}{@{}p{2.8cm}|ccccl} 


circuit diagram 

&

\input{pbcircuit-NotXAndY.tex} 

& $=$ &

\input{pbcircuit-XAndY.tex}

& $\compose$ & 

\input{pbcircuit-NotWire.tex}

\\

output formulae 
& ($\logicneg{x} \logicand y$) & $=$ & ($x \logicand y$) & $\compose$ & ($\logicneg{x}$, $y$) \\

$(m,n) =$ (\#outputs, \#inputs) & $(1,2)$ & &  $(1,2)$ & & $(2,2)$ \\

\hline

fourier matrix dims ($2^m \times 2^n$) & $2 \times 4$ & &  $2 \times 4$ & & $4 \times 4$ \\ 

fourier matrix (column order is $1$, $y$, $x$, $xy$)
 
& 
$\begin{vmatrix} 1 & 0 & 0 & 0 \\ \frac{1}{2} & \frac{1}{2} & $-$ \frac{1}{2} & \frac{1}{2} \end{vmatrix}$ 
& $=$ 
&  
$\begin{vmatrix} 1 & 0 & 0 & 0 \\ \frac{1}{2} & \frac{1}{2} & \frac{1}{2} & $-$ \frac{1}{2} \end{vmatrix}$
& $\times$ 
& 
$\begin{vmatrix} 1 & 0 & 0 & 0 \\ 0 & $1$ & 0 & 0 \\ 0 & 0 & $-1$ & 0 \\ 0 & 0 & 0 & $-1$ \end{vmatrix}$ 
\\ 

\hline

pattern matrix dims ($2^{2^m} \times 2^{2^n}$) & $4 \times 16$ & &  $4 \times 16$ & & $16 \times 16$ \\

 & & & & &
\multirow{9}{*}{
{ \tiny \arraycolsep=1.4pt
$
\begin{vmatrix}
1  &  0  &  0  &  0  &  0  &  0  &  0  &  0  &  0  &  0  &  0  &  0  &  0  &  0  &  0  &  0  \\
0  &  0  &  0  &  0  &  1  &  0  &  0  &  0  &  0  &  0  &  0  &  0  &  0  &  0  &  0  &  0  \\
0  &  0  &  0  &  0  &  0  &  0  &  0  &  1  &  0  &  0  &  0  &  0  &  0  &  0  &  0  &  0  \\
0  &  0  &  0  &  1  &  0  &  0  &  0  &  0  &  0  &  0  &  0  &  0  &  0  &  0  &  0  &  0  \\
0  &  1  &  0  &  0  &  0  &  0  &  0  &  0  &  0  &  0  &  0  &  0  &  0  &  0  &  0  &  0  \\
0  &  0  &  0  &  0  &  0  &  1  &  0  &  0  &  0  &  0  &  0  &  0  &  0  &  0  &  0  &  0  \\
0  &  0  &  0  &  0  &  0  &  0  &  1  &  0  &  0  &  0  &  0  &  0  &  0  &  0  &  0  &  0  \\
0  &  0  &  1  &  0  &  0  &  0  &  0  &  0  &  0  &  0  &  0  &  0  &  0  &  0  &  0  &  0  \\
0  &  0  &  0  &  0  &  0  &  0  &  0  &  0  &  1  &  0  &  0  &  0  &  0  &  0  &  0  &  0  \\
0  &  0  &  0  &  0  &  0  &  0  &  0  &  0  &  0  &  - \! 1  &  0  &  0  &  0  &  0  &  0  &  0  \\
0  &  0  &  0  &  0  &  0  &  0  &  0  &  0  &  0  &  0  &  1  &  0  &  0  &  0  &  0  &  0  \\
0  &  0  &  0  &  0  &  0  &  0  &  0  &  0  &  0  &  0  &  0  &  - \! 1  &  0  &  0  &  0  &  0  \\
0  &  0  &  0  &  0  &  0  &  0  &  0  &  0  &  0  &  0  &  0  &  0  &  - \! 1  &  0  &  0  &  0  \\
0  &  0  &  0  &  0  &  0  &  0  &  0  &  0  &  0  &  0  &  0  &  0  &  0  &  1  &  0  &  0  \\
0  &  0  &  0  &  0  &  0  &  0  &  0  &  0  &  0  &  0  &  0  &  0  &  0  &  0  &  - \! 1  &  0  \\
0  &  0  &  0  &  0  &  0  &  0  &  0  &  0  &  0  &  0  &  0  &  0  &  0  &  0  &  0  &  1
\end{vmatrix}
$
}
}

\\

\multirow{3}{2.8cm}{pattern matrix ($\PM_{\P}$),
using a pattern basis $\P = \P_{\RmTwob}$
described below,
defined for
$m, n \in \{0,1,2\}$
}

& 
\multicolumn{3}{l}{ 
	{ \tiny \arraycolsep=1.4pt
	$
\begin{vmatrix}
1  &  0  &  0  &  0  &  0  &  0  &  0  &  0  &  0  &  0  &  0  &  0  &  0  &  0  &  0  &  0  \\
0  &  \frac{1}{2}  &  \frac{1}{2}  &  0  &  - \! \frac{1}{2}  &  0  &  0  &  \frac{1}{2}  &  0  &  0  &  0  &  0  &  0  &  0  &  0  &  0  \\
0  &  0  &  0  &  0  &  0  &  0  &  0  &  0  &  \frac{1}{2}  &  0  &  0  &  - \! \frac{1}{2}  &  0  &  \frac{1}{2}  &  \frac{1}{2}  &  0  \\
0  &  0  &  0  &  0  &  0  &  0  &  0  &  0  &  0  &  0  &  0  &  0  &  0  &  0  &  0  &  1
\end{vmatrix}
	$
	}
=}
& 
& 
\\ 
\\
&
\multicolumn{3}{r}{ 
	{ \tiny \arraycolsep=1.4pt
	$
\begin{vmatrix}
1  &  0  &  0  &  0  &  0  &  0  &  0  &  0  &  0  &  0  &  0  &  0  &  0  &  0  &  0  &  0  \\
0  &  - \! \frac{1}{2}  &  \frac{1}{2}  &  0  &  \frac{1}{2}  &  0  &  0  &  \frac{1}{2}  &  0  &  0  &  0  &  0  &  0  &  0  &  0  &  0  \\
0  &  0  &  0  &  0  &  0  &  0  &  0  &  0  &  \frac{1}{2}  &  0  &  0  &  \frac{1}{2}  &  0  &  \frac{1}{2}  &  - \! \frac{1}{2}  &  0  \\
0  &  0  &  0  &  0  &  0  &  0  &  0  &  0  &  0  &  0  &  0  &  0  &  0  &  0  &  0  &  1
\end{vmatrix}
	$
	}
}
& $\times$  &  
\\

\\
computation of two variants of measured complexity $C_{M,\P}(f)$: \\

$M_{\nz}$ of matrix 
&
4 \quad\quad\quad &
$\le$&
\quad\quad\quad 4 &
$\times$&
\quad\, 1 \\

$C_{M,\P} =
\log_2 M_{\nz} \! $
&
2 \quad\quad\quad &
$\le$&
\quad\quad\quad 2 &
$+$&
\quad\, 0 \\

$M_{\abs}$ of matrix 
&
2 \quad\quad\quad &
$\le$&
\quad\quad\quad 2 &
$\times$&
\quad\, 1 \\

$C_{M,\P} =
\log_2 M_{\abs} \! \! \! $
&
1 \quad\quad\quad &
$\le$&
\quad\quad\quad 1 &
$+$&
\quad\, 0 \\

\end{tabular} 

}

\newpage

\subsubsection{examples of pattern bases\label{S:egpatbases}} 

Here are the
change-of-basis matrices
which define several pattern bases.
(The details of this representation are described in a later subsection.)
The first two are defined for all $n$;
the others are defined only for the values of $n$ shown here.

\bigskip
\begin{itemize}
\item
The indicator basis $\P_{\I}$
and the monomial basis $\P_{\mon}$
were described earlier in the Summary, and are easy to define for all $n$ 
(using identity matrices, or discrete fourier matrices over the group $\F_2^{2^n}$, respectively):
{ \tiny \arraycolsep=2.8pt
\begin{equation*}
P_0^{(\I)} =
\begin{vmatrix}
1 & 0 \\
0 & 1
\end{vmatrix}
,
\quad
P_1^{(\I)} =
\begin{vmatrix}
1 & 0 & 0 & 0 \\
0 & 1 & 0 & 0 \\
0 & 0 & 1 & 0 \\
0 & 0 & 0 & 1
\end{vmatrix}
,
\quad
\dots
\end{equation*}
}
{ \tiny \arraycolsep=2.8pt
\begin{equation*}
P_0^{(\mon)} =
\tfrac{1}{\sqrt{2}}
\begin{vmatrix}
1 & 1 \\
1 & -1
\end{vmatrix}
,
\quad
P_1^{(\mon)} =
\frac{1}{2}
\begin{vmatrix}
1 &  1 &  1 &  1 \\
1 & -1 &  1 & -1 \\
1 &  1 & -1 & -1 \\
1 & -1 & -1 &  1
\end{vmatrix}
,
\quad
\dots
\end{equation*}
}
\item
The basis 
$\P_{\RmTwob}$
was discovered as part of a classification (to be published separately)
of all ``nice'' real pattern bases defined for $n \in \{0,1,2\}$,
where ``nice'' means orthonormal (in the canonical basis of each property space)
and giving all $\F_2$-affine functions
a measured complexity of zero.
It was used to compute the pattern matrices shown in 
the table above.
{ \tiny \arraycolsep=2.8pt
\begin{equation*}
P_0^{(\RmTwob)} =
\tfrac{1}{\sqrt{2}}
\begin{vmatrix}
1 & 1 \\
1 & -1
\end{vmatrix}
,
\quad
P_1^{(\RmTwob)} =
\frac{1}{\sqrt{2}}
\begin{vmatrix}
1 & 0 & 0 & 1 \\
0 & 1 & 1 & 0 \\
0 & 1 & -1 & 0 \\
1 & 0 & 0 & -1
\end{vmatrix}
,
\end{equation*}
}
{ \tiny \arraycolsep=1.4pt
\begin{equation*}
P_2^{(\RmTwob)} = \frac{1}{2 \sqrt{2}} 
\begin{vmatrix}
2  &  0  &  0  &  0  &  0  &  0  &  0  &  0  &  0  &  0  &  0  &  0  &  0  &  0  &  0  &  2  \\
0  &  - \! 1  &  1  &  0  &  1  &  0  &  0  &  1  &  1  &  0  &  0  &  1  &  0  &  1  &  - \! 1  &  0  \\
0  &  1  &  - \! 1  &  0  &  1  &  0  &  0  &  1  &  1  &  0  &  0  &  1  &  0  &  - \! 1  &  1  &  0  \\
0  &  0  &  0  &  2  &  0  &  0  &  0  &  0  &  0  &  0  &  0  &  0  &  2  &  0  &  0  &  0  \\
0  &  1  &  1  &  0  &  - \! 1  &  0  &  0  &  1  &  1  &  0  &  0  &  - \! 1  &  0  &  1  &  1  &  0  \\
0  &  0  &  0  &  0  &  0  &  2  &  0  &  0  &  0  &  0  &  2  &  0  &  0  &  0  &  0  &  0  \\
0  &  0  &  0  &  0  &  0  &  0  &  2  &  0  &  0  &  2  &  0  &  0  &  0  &  0  &  0  &  0  \\
0  &  1  &  1  &  0  &  1  &  0  &  0  &  - \! 1  &  - \! 1  &  0  &  0  &  1  &  0  &  1  &  1  &  0  \\
0  &  1  &  1  &  0  &  1  &  0  &  0  &  - \! 1  &  1  &  0  &  0  &  - \! 1  &  0  &  - \! 1  &  - \! 1  &  0  \\
0  &  0  &  0  &  0  &  0  &  0  &  2  &  0  &  0  &  - \! 2  &  0  &  0  &  0  &  0  &  0  &  0  \\
0  &  0  &  0  &  0  &  0  &  2  &  0  &  0  &  0  &  0  &  - \! 2  &  0  &  0  &  0  &  0  &  0  \\
0  &  1  &  1  &  0  &  - \! 1  &  0  &  0  &  1  &  - \! 1  &  0  &  0  &  1  &  0  &  - \! 1  &  - \! 1  &  0  \\
0  &  0  &  0  &  2  &  0  &  0  &  0  &  0  &  0  &  0  &  0  &  0  &  - \! 2  &  0  &  0  &  0  \\
0  &  1  &  - \! 1  &  0  &  1  &  0  &  0  &  1  &  - \! 1  &  0  &  0  &  - \! 1  &  0  &  1  &  - \! 1  &  0  \\
0  &  - \! 1  &  1  &  0  &  1  &  0  &  0  &  1  &  - \! 1  &  0  &  0  &  - \! 1  &  0  &  - \! 1  &  1  &  0  \\
2  &  0  &  0  &  0  &  0  &  0  &  0  &  0  &  0  &  0  &  0  &  0  &  0  &  0  &  0  &  - \! 2
\end{vmatrix}
\end{equation*}
}

(For an explanation of 
$P_2^{(\RmTwob)}$'s
structure, 
and for more example pattern bases (some definable for all~$n$),
see the separate classification mentioned above.)

\item
If we allow matrix entries in $\C$, 
there is 
another
``nice'' basis defined for $n \in \{0,1\}$,
which I call $\P_{\COne}$. 
(It can be extended in at least one way to a nice but useless basis for all $n$, but 
I don't yet know whether it can be extended usefully to $n=2$.
In this context, ``useful'' just means ``nice, but giving nonzero measured complexity to AND'',
since the ``adjoined wire issue'' can't be studied for nontrivial functions until $n=3$.)
{ \small \arraycolsep=2.8pt
\begin{equation*}
P_0^{(\COne)} =
\tfrac{1}{\sqrt{2}}
\begin{vmatrix}
1 & -i \\
1 & i
\end{vmatrix}
,
\quad
P_1^{(\COne)} =
\frac{1}{\sqrt{2}}
\begin{vmatrix}
1 & 0 & 0 & -i \\
0 & 1 & -i & 0 \\
0 & 1 & i & 0 \\
1 & 0 & 0 & i
\end{vmatrix}
\end{equation*}
}
\item
Finally, 
Table~2 (which shows measured complexities for various examples of pattern basis)
includes 
the following randomly chosen real orthonormal pattern basis,\footnote{
    When using $\P_{\randOne}$ in making Table~2, 
    its change of basis matrices were ``re-orthonormalized'' 
    (to undo the effect of rounding when printing at limited precision), by subtracting
    from each row some multiple of each prior row (to make the matrix orthogonal), then normalizing that row.
    In principle, this is not needed, since even a set of non-orthogonal change of basis matrices should induce
    a ``valid'' (that is, subadditive) measured complexity function.
}
$\P_{\randOne}$,
mainly to show how comparatively badly it behaves,
but also to demonstrate that the resulting measured complexity is still subadditive: 
{ \tiny \arraycolsep=2.8pt
\begin{equation*}
P_0^{(\randOne)} =
\begin{vmatrix}
0.7423  &  0.6701  \\
0.6701  &  - \! 0.7423
\end{vmatrix}
,
\quad
P_1^{(\randOne)} =
\begin{vmatrix}
- \! 0.1861  &  0.2500  &  0.9456  &  0.0932  \\
- \! 0.7947  &  - \! 0.5938  &  0.0130  &  - \! 0.1255  \\
- \! 0.1048  &  0.3376  &  - \! 0.0177  &  - \! 0.9353  \\
- \! 0.5682  &  0.6863  &  - \! 0.3246  &  0.3175
\end{vmatrix}
,
\quad
P_2^{(\randOne)} =
\end{equation*}
}
{ \tiny \arraycolsep=1.4pt
\begin{equation*}
\begin{vmatrix}
0.0401  &  0.1033  &  0.4851  &  0.0579  &  0.0534  &  - \! 0.0545  &  0.2001  &  - \! 0.2051  &  - \! 0.2003  &  0.1588  &  0.1461  &  - \! 0.2735  &  - \! 0.0270  &  0.0970  &  - \! 0.4995  &  0.4899  \\
0.1799  &  0.1515  &  - \! 0.0305  &  0.3457  &  0.2003  &  0.1240  &  - \! 0.1366  &  - \! 0.1682  &  0.0574  &  0.4958  &  0.1048  &  - \! 0.2710  &  0.2377  &  - \! 0.4962  &  - \! 0.0040  &  - \! 0.2926  \\
- \! 0.2397  &  0.2374  &  - \! 0.1010  &  - \! 0.2713  &  - \! 0.1289  &  0.3596  &  0.5107  &  0.1346  &  0.1842  &  - \! 0.0054  &  0.1301  &  - \! 0.0524  &  0.5524  &  - \! 0.0720  &  0.0479  &  0.1060  \\
- \! 0.2206  &  - \! 0.2561  &  - \! 0.1878  &  - \! 0.0913  &  - \! 0.0906  &  0.3089  &  0.1046  &  - \! 0.7256  &  - \! 0.0383  &  0.1245  &  0.1685  &  0.1292  &  - \! 0.1349  &  0.2131  &  - \! 0.1247  &  - \! 0.2445  \\
- \! 0.3093  &  0.2237  &  0.0215  &  - \! 0.0378  &  - \! 0.3879  &  - \! 0.1652  &  - \! 0.2471  &  0.1571  &  0.3165  &  - \! 0.0378  &  0.4255  &  0.1125  &  - \! 0.1906  &  - \! 0.2128  &  - \! 0.4264  &  - \! 0.1736  \\
0.3320  &  - \! 0.0218  &  - \! 0.2168  &  - \! 0.5501  &  0.1374  &  - \! 0.3821  &  - \! 0.2969  &  - \! 0.2484  &  0.1021  &  0.0556  &  0.1572  &  0.0973  &  0.3706  &  - \! 0.0093  &  - \! 0.0909  &  0.1777  \\
0.3285  &  0.0175  &  - \! 0.1584  &  0.0547  &  0.2245  &  0.1865  &  0.0326  &  0.0194  &  0.3511  &  - \! 0.3985  &  0.3868  &  - \! 0.5010  &  - \! 0.1901  &  0.2343  &  0.0110  &  - \! 0.0572  \\
- \! 0.0340  &  - \! 0.1974  &  - \! 0.5539  &  0.2068  &  - \! 0.3774  &  - \! 0.3196  &  0.2262  &  - \! 0.0111  &  - \! 0.0194  &  0.2083  &  0.0616  &  - \! 0.2733  &  - \! 0.1458  &  - \! 0.1100  &  0.1300  &  0.3768  \\
- \! 0.3937  &  0.1912  &  0.0712  &  0.1167  &  - \! 0.0704  &  0.0816  &  - \! 0.5090  &  - \! 0.1852  &  0.3240  &  0.0877  &  - \! 0.2200  &  - \! 0.2549  &  0.1478  &  0.3067  &  0.2495  &  0.2839  \\
- \! 0.3651  &  - \! 0.2046  &  0.3340  &  - \! 0.2909  &  0.2958  &  - \! 0.3435  &  0.2287  &  - \! 0.1100  &  0.2126  &  0.0340  &  0.1068  &  - \! 0.1966  &  - \! 0.2108  &  - \! 0.2773  &  0.3665  &  - \! 0.0820  \\
- \! 0.1158  &  - \! 0.3351  &  - \! 0.1537  &  - \! 0.1068  &  0.2202  &  - \! 0.0429  &  0.0851  &  0.2866  &  0.3236  &  0.2979  &  - \! 0.4000  &  - \! 0.1700  &  0.0213  &  0.2113  &  - \! 0.4900  &  - \! 0.1787  \\
- \! 0.1534  &  - \! 0.1713  &  0.0969  &  - \! 0.0191  &  - \! 0.1826  &  - \! 0.2198  &  - \! 0.1115  &  0.1779  &  - \! 0.4626  &  0.0266  &  0.2690  &  - \! 0.3673  &  0.3446  &  0.3388  &  0.0705  &  - \! 0.3888  \\
- \! 0.0047  &  - \! 0.6955  &  0.1873  &  0.1598  &  - \! 0.1443  &  0.2070  &  - \! 0.1961  &  0.0064  &  0.1120  &  - \! 0.3116  &  0.0483  &  0.0035  &  0.2969  &  - \! 0.3298  &  - \! 0.0858  &  0.1927  \\
0.4036  &  - \! 0.1441  &  0.3183  &  - \! 0.3010  &  - \! 0.4539  &  0.2433  &  - \! 0.0313  &  0.1434  &  0.1884  &  0.4376  &  0.0412  &  - \! 0.0331  &  - \! 0.1965  &  0.1119  &  0.2373  &  - \! 0.0195  \\
0.1226  &  - \! 0.0723  &  0.1774  &  0.4682  &  0.0238  &  - \! 0.3267  &  0.2514  &  - \! 0.0513  &  0.3845  &  0.1070  &  0.2085  &  0.3956  &  0.2522  &  0.3448  &  0.1064  &  - \! 0.0536  \\
- \! 0.2043  &  - \! 0.1577  &  - \! 0.1465  &  0.0018  &  0.4103  &  0.2512  &  - \! 0.1856  &  0.3385  &  - \! 0.1491  &  0.3292  &  0.4727  &  0.2439  &  - \! 0.1115  &  0.1043  &  0.0945  &  0.2926
\end{vmatrix}
\end{equation*}
}
\end{itemize}

\newpage

\subsubsection{pattern matrix for the AND function\label{S:patmatforAND}}

As a detailed example of how to construct a pattern matrix, 
we'll study the 2-input AND function (called $x \logicand y$ in Table~1),
which maps $(x,y) \in \F_2^2$ to $xy \in \F_2$.
(This function 
has $n = 2$ inputs and $m = 1$ output, 
so its pattern matrix dimensions are 
$2^{2^m} = 2^{2^1} = 4$
by 
$2^{2^n} = 2^{2^2} = 16$.)

\bigskip
We'll start with the simplest possible pattern basis, $\P_{\I}$; 
this uses the indicator basis,
which is also the canonical basis, in each property space.
This will give us a trivial pattern matrix (shown below in~\eqref{E:PMindAND}),
and a measured complexity of 0 for every function,
but it makes the pattern matrix easy to understand
(and its basis can be changed later).

Using this basis,
each row of the pattern matrix $\PM_{\P_{\I}}(f)$
corresponds to an $m$-input 1-output truth table (or an ``$(m,1)$-table'' for short).
We'll order the rows lexicographically, so for $m=1$ they are labelled by:
the constant function $0$ (with one input, which it ignores), 
the identity function (sometimes called a ``wire''),
the NOT function, 
and the constant function $1$.
(For convenience, we'll often identify a truth table with the function it describes.)

Each column of this pattern matrix 
corresponds to one $n$-input 1-output truth table
(or ``$(n,1)$-table''),
of which there are $2^{2^n} = 16$.
We'll put these in ``doubly-lexicographic order''~---
the possible values of the 2 inputs (x,y) 
will be ordered $(0,0)$, $(0,1)$, $(1,0)$, $(1,1)$,
which tells us how to arrange each truth table's 4 output bits as a 4-tuple;
the 16 possible 4-tuples will be sorted lexicographically
to label the columns,
giving a column order of 
$(0,0,0,0)$,
$(0,0,0,1)$,
$(0,0,1,0)$,
$(0,0,1,1)$,
\dots
$(1,1,1,0)$,
$(1,1,1,1)$.
(The AND function itself has a truth table of $(0,0,0,1)$ in this format, so it
labels the 2nd column in this order.)

To construct the indicator-basis pattern matrix for AND (or for any function $f$),
we compose its output with each $(m,1)$-table $t_o$ in turn,
and see which $(n,1)$-table $t_i$ is induced on its inputs~---
i.e., which $(n,1)$-table describes the composition $t_o \compose f$.
This tells us which matrix entry is 1 (with all others being 0) 
in the row labelled by $t_o$.
(This is a kind of ``pullback'' operation on $f$, 
as we'll discuss when the pattern matrix is defined formally in Section~\ref{S:DefPatternMatrix}.)

In the special case of $m = 1$,
the 2nd row is labelled by $t_o = (0,1)$ (which describes the identity function on $\F_2$),
so that row just indicates $f$'s ordinary truth table
(by putting its single 1 in the corresponding column).
For the AND function, whose truth table is $(0,0,0,1)$,
that row gets a 1 in the 2nd column
(as can be seen in the full matrix shown below).

The 3rd row is labelled by $t_o = (1,0)$ (the NOT function),
so it just indicates the truth table of NOT($f$) (in the same manner).
The 1st and 4th rows (labelled by a $t_o$ of $(0,0)$ and $(1,1)$ respectively, which describe constant functions)
are the same for all functions (provided we're using this basis $\P_{\I}$ and $m = 1$);
they indicate the all-0 and all-1 $n$-input truth tables respectively,
by placing their single 1 entry in the first and last columns respectively.

\bigskip
For the AND function, 
we end up with
the following indicator-basis pattern matrix 
(which differs from AND's pattern matrix in Table~1,
because we used a different pattern basis there):
\begin{equation}\label{E:PMindAND}
A \,=\, \PM_{\P_{\I}}(\text{AND}) \,=\,
\begin{vmatrix}
1 & 0 & 0 & 0 & 0 & 0 & 0 & 0 & 0 & 0 & 0 & 0 & 0 & 0 & 0 & 0 \\ 
0 & 1 & 0 & 0 & 0 & 0 & 0 & 0 & 0 & 0 & 0 & 0 & 0 & 0 & 0 & 0 \\ 
0 & 0 & 0 & 0 & 0 & 0 & 0 & 0 & 0 & 0 & 0 & 0 & 0 & 0 & 1 & 0 \\ 
0 & 0 & 0 & 0 & 0 & 0 & 0 & 0 & 0 & 0 & 0 & 0 & 0 & 0 & 0 & 1
\end{vmatrix}
\end{equation}

Of course this pattern matrix is not a very {\it compact} encoding of the AND function!
But that's not its purpose~---
what interests us is 
(1) its matrix multiplication corresponds to function composition;
(2) after suitable changes of basis, 
its entries can tell us the relationship between {\it any} properties
of $t_o$
and $t_i$,
in any circuit which computes $t_i$ using the structure $t_o \compose f$~---
that is, any circuit whose first stage is $f$ and whose remainder is described by $t_o$.
This lets the pattern matrix
{\it linearly} encode everything about $f$'s possible role in any 1-output circuit's computation
(provided $f$ has been ``widened'' to include whatever wires run alongside it
in the circuit, as described earlier).
And feature (1) makes it possible
to use this matrix (in {\it any} basis)
to define
subadditive measures with respect to function composition
(which any generally useful circuit complexity lower bound ought to be).

(The fourier matrices mentioned earlier also
have feature (1) from that list~--- 
indeed, they {\it can} be used to define subadditive measures on boolean functions~---
but they lack feature (2),
which is part of the reason
the subadditive measures they help define can never have nontrivially large values.
A more basic reason is just their smaller size, 
though this could be seen as a different aspect of the same limitation.)

\bigskip
Of course, none of this will be interesting until we {\it do}
change that matrix's basis
(which requires {\it finding} a useful basis to change it to, as we've discussed elsewhere).
For now,
we'll just show how we got the pattern matrices in Table~1,
by changing from the indicator basis to $\P_{\RmTwob}$,
which is ``useful'' for $n=2$.
(Whether $\P_{\RmTwob}$ is usefully extendable to higher $n$ is not yet known.)

\subsubsection{change of pattern basis\label{S:changeofpatternbasis}}

First we should describe our conventions for change of basis matrices (which follow \cite{MB99}).
A change of basis matrix $P$
should consist of columns giving each ``old'' basis element
in ``new'' coordinates (i.e. in coordinates relative to the new basis).
A column vector $v$, using old coordinates,
can then be changed to a vector $v'$, using new coordinates, by
$$
v' = A \, v
$$
and a square matrix $A$, represented in the old basis,
can be changed to $A'$, expressing the same linear map relative to the new basis,
using
$$
A' = P \, A \, P^{-1}
$$

Since a pattern matrix is in general rectangular, 
we need to use families of bases (one for each possible matrix dimension).
By an abuse of notation, within a pattern basis $\P$ we'll let $P_n = \P_{P_n}$ (for each $n \ge 0$) denote 
not only a basis set $P_n \subset Q_n$
(of the property space $Q_n = \P_{Q_n}$),
but also the change of basis matrix from the indicator basis to that basis set
(ordering its rows and columns in a standard way implied by the context).\footnote{
    This overloaded meaning for the notation $P_n$
    is compatible with ``tensoring basis sets''
    (i.e., forming a basis for a tensor product of vector spaces $V_1 \tensor V_2$,
     by tensoring every pair of elements of existing bases for $V_1$ and $V_2$)~---
    for example,
    $P_n \tensor P_n$ 
    has a compatible meaning,
    whether we interpret $P_n$ as a basis set,
    or as that basis set's change of basis matrix from the canonical (indicator) basis.
    This also explains the notation $P_{n+1} = W_{n+1} \big( P_n \tensor P_n \big)$
    for the ``appropriate change of basis'' (in an inductive construction of the basis family $\langle P_n \rangle$) 
    which we discussed
    in Section~\ref{S:tensorwithitself}.
}
If $A$ is a pattern matrix representing the pattern map of $f \colon \F_2^n \to \F_2^m$ using the indicator pattern basis
(as in~\eqref{E:PMindAND}),
we can change it to a different pattern matrix $A'$, 
representing the same pattern map
using a different pattern basis $\P$ 
(containing the property space bases (and change of basis matrices) $\langle P_n \rangle$), by
\begin{equation*}
A' = P_m \, A \, P_n^{-1}
\end{equation*}

\bigskip
In the notation we used earlier, in which the pattern basis used by a pattern matrix is indicated by a subscript on $\PM$
(so the pattern matrices $\PM_{\P}(f)$ and $\PM_{\P_{\I}}(f)$ 
use the pattern bases $\P$ and $\P_{\I}$ respectively),
we can express that instead as
\begin{equation} 
\PM_{\P}(f) = P_m \,\, \PM_{\P_{\I}}(f) \,\, P_n^{-1}
\end{equation}

Thus,
to construct $\PM_{\P_{\RmTwob}}(\text{AND})$ as shown in Table~1,
we have:
\begin{align*}
&\PM_{\P_{\RmTwob}}(\text{AND}) 
\,\, = \,\, P_1^{(\RmTwob)} \,\,\, \PM_{\P_{\I}}(\text{AND}) \,\,\, P_2^{-1(\RmTwob)} \notag \\
&=
{ \tiny \arraycolsep=1.4pt
\tfrac{1}{\sqrt{2}}
\begin{vmatrix}
1 & 0 & 0 & 1 \\
0 & 1 & 1 & 0 \\
0 & 1 & -1 & 0 \\
1 & 0 & 0 & -1
\end{vmatrix}
\,\times\,
\begin{vmatrix}
1 & 0 & 0 & 0 & 0 & 0 & 0 & 0 & 0 & 0 & 0 & 0 & 0 & 0 & 0 & 0 \\ 
0 & 1 & 0 & 0 & 0 & 0 & 0 & 0 & 0 & 0 & 0 & 0 & 0 & 0 & 0 & 0 \\ 
0 & 0 & 0 & 0 & 0 & 0 & 0 & 0 & 0 & 0 & 0 & 0 & 0 & 0 & 1 & 0 \\ 
0 & 0 & 0 & 0 & 0 & 0 & 0 & 0 & 0 & 0 & 0 & 0 & 0 & 0 & 0 & 1
\end{vmatrix}
\,\times\,
\frac{1}{2 \sqrt{2}} 
\begin{vmatrix}
2  &  0  &  0  &  0  &  0  &  0  &  0  &  0  &  0  &  0  &  0  &  0  &  0  &  0  &  0  &  2  \\
0  &  - \! 1  &  1  &  0  &  1  &  0  &  0  &  1  &  1  &  0  &  0  &  1  &  0  &  1  &  - \! 1  &  0  \\
0  &  1  &  - \! 1  &  0  &  1  &  0  &  0  &  1  &  1  &  0  &  0  &  1  &  0  &  - \! 1  &  1  &  0  \\
0  &  0  &  0  &  2  &  0  &  0  &  0  &  0  &  0  &  0  &  0  &  0  &  2  &  0  &  0  &  0  \\
0  &  1  &  1  &  0  &  - \! 1  &  0  &  0  &  1  &  1  &  0  &  0  &  - \! 1  &  0  &  1  &  1  &  0  \\
0  &  0  &  0  &  0  &  0  &  2  &  0  &  0  &  0  &  0  &  2  &  0  &  0  &  0  &  0  &  0  \\
0  &  0  &  0  &  0  &  0  &  0  &  2  &  0  &  0  &  2  &  0  &  0  &  0  &  0  &  0  &  0  \\
0  &  1  &  1  &  0  &  1  &  0  &  0  &  - \! 1  &  - \! 1  &  0  &  0  &  1  &  0  &  1  &  1  &  0  \\
0  &  1  &  1  &  0  &  1  &  0  &  0  &  - \! 1  &  1  &  0  &  0  &  - \! 1  &  0  &  - \! 1  &  - \! 1  &  0  \\
0  &  0  &  0  &  0  &  0  &  0  &  2  &  0  &  0  &  - \! 2  &  0  &  0  &  0  &  0  &  0  &  0  \\
0  &  0  &  0  &  0  &  0  &  2  &  0  &  0  &  0  &  0  &  - \! 2  &  0  &  0  &  0  &  0  &  0  \\
0  &  1  &  1  &  0  &  - \! 1  &  0  &  0  &  1  &  - \! 1  &  0  &  0  &  1  &  0  &  - \! 1  &  - \! 1  &  0  \\
0  &  0  &  0  &  2  &  0  &  0  &  0  &  0  &  0  &  0  &  0  &  0  &  - \! 2  &  0  &  0  &  0  \\
0  &  1  &  - \! 1  &  0  &  1  &  0  &  0  &  1  &  - \! 1  &  0  &  0  &  - \! 1  &  0  &  1  &  - \! 1  &  0  \\
0  &  - \! 1  &  1  &  0  &  1  &  0  &  0  &  1  &  - \! 1  &  0  &  0  &  - \! 1  &  0  &  - \! 1  &  1  &  0  \\
2  &  0  &  0  &  0  &  0  &  0  &  0  &  0  &  0  &  0  &  0  &  0  &  0  &  0  &  0  &  - \! 2
\end{vmatrix}
}
\\
&=
{ \tiny \arraycolsep=1.4pt
\begin{vmatrix}
1  &  0  &  0  &  0  &  0  &  0  &  0  &  0  &  0  &  0  &  0  &  0  &  0  &  0  &  0  &  0  \\
0  &  - \! \frac{1}{2}  &  \frac{1}{2}  &  0  &  \frac{1}{2}  &  0  &  0  &  \frac{1}{2}  &  0  &  0  &  0  &  0  &  0  &  0  &  0  &  0  \\
0  &  0  &  0  &  0  &  0  &  0  &  0  &  0  &  \frac{1}{2}  &  0  &  0  &  \frac{1}{2}  &  0  &  \frac{1}{2}  &  - \! \frac{1}{2}  &  0  \\
0  &  0  &  0  &  0  &  0  &  0  &  0  &  0  &  0  &  0  &  0  &  0  &  0  &  0  &  0  &  1
\end{vmatrix}
}
\end{align*}

\bigskip
(Note that $P_2^{-1(\RmTwob)} = P_2^{(\RmTwob)}$, since it's orthonormal and symmetric.\footnote{
    In general, a pattern basis change of basis matrix 
    is not necessarily even orthogonal (or unitary), let alone also symmetric;
    even when it can be expressed that way, this
    depends on the arbitrary order and sign (or complex phase) of the basis elements.
    But in all our listed examples (except $\P_{\randOne}$),
    it's either already true, or can be made true by rescaling some patterns by roots of unity.
})

\subsubsection{comparing choices of pattern basis $\P$ and matrix measure $M$\label{S:comparingchoices}}

Finally, we
compare the measured complexity
given by
several choices of pattern basis $\P$ and matrix measure $M$,
noting that it's always subadditive (with respect to function composition),
even though it's not always useful,
or related to actual circuit complexity.

In the following table, we compare several variants of $C_{M,\P}$ (and two kinds of actual complexity),
for a few simple functions
with 1 or 2 outputs and 0 to 2 inputs,
including the functions that were shown in Table~1.
In the first three entries of each numeric column, 
the reader can verify subadditivity of that column's
complexity measure,
since the first row's function ($\logicneg{x} \logicand y$)
is the composition of the functions in the next two rows (($x \logicand y$) and ($\logicneg{x}$, $y$)).
Other observations about the tabulated values are discussed below.

\begin{center}
\textbf{Table 2:} value of $C_{M,\P}(f)$, for various functions $f$ and choices of $M$ and $\P$, \\
compared to two kinds of actual complexity
\end{center}

\noindent\begin{tabular}{p{2.0cm}|p{2.0cm}|cc|cc|cc||cc|}

&
choice of $\P$:\! &
\multicolumn{2}{c|}{ $\P_{\mon}$ } & 
\multicolumn{2}{c|}{ $\P_{\RmTwob}$ } & 
\multicolumn{2}{c||}{ $\P_{\randOne}$ } & 
\multicolumn{2}{p{3cm}|}{ \parbox[t]{3.3cm}{or, kind of \\ actual~complexity:} } \\ 

&
choice of $M\!$:\!\! &
$M_{\nz}$ & $M_{\abs}$ & 
$M_{\nz}$ & $M_{\abs}$ & 
$M_{\nz}$ & $M_{\abs}$ & 
$C_{B_2}$ & ``free XORs''  \\

\hline

\parbox[t]{2.0cm}{
2-input functions: 
} & 
(function nickname) & && && && & \\

($\logicneg{x} \logicand y$)  & 
AND($\logicneg{x}$, $y$) & 
2 & 1 &
2 & 1 &
4 & 1.807 &
1 & 1  \\

($x \logicand y$) & 
AND & 
2 & 1 &
2 & 1 &
4 & 1.840 &
1 & 1  \\

($\logicneg{x}$, $y$) & 
{\it complement first input} & 
0 & 0 &
0 & 0 &
4 & 1.812 &
0 & 0  \\

($x \logicxor y$, $y$) & 
{\it controlled-}NOT & 
0 & 0 &
0 & 0 &
4 & 1.772 &
1 & 0  \\

($x \logicxor y$)  & 
XOR & 
2 & 1 &
0 & 0 &
4 & 1.834 &
1 & 0  \\

($x$)
 & 
{\it discard 2nd input} & 
2 & 1 &
0 & 0 &
4 & 1.846 &
0 & 0  \\

\hline

\parbox[t]{2.0cm}{
1-input $f$s:
} & & && && && & \\

($x$, $x$)  & 
{\it splitter} & 
0 & 1 &
0 & 0 &
2 & 1.385 &
0 & 0  \\

($\logicneg{x}$)  & 
NOT & 
0 & 0 &
0 & 0 &
2 & 0.917 &
0 & 0  \\

\hline

\parbox[t]{2.0cm}{
0-input $f$s:
} & & && && && & \\

($0$)  & 
{\it create 0} & 
0 & $\frac{1}{2}$ &
0 & 0 &
1 & 0.966 &
0 & 0  \\

\end{tabular}

\bigskip
The functions are represented 
as tuples of output formulae (as in Table~1),
in which the boolean operations AND, XOR, and NOT
are represented by $\logicand$, $\logicxor$, and $\logicneg{x}$
respectively. 
The 0-input function shown as~$(0)$ has one output, which is always $0 \in \F_2$.
The pattern bases $\P_{\dots}$ are defined in Section~\ref{S:egpatbases}; 
we don't show $\P_{\I}$ here, since its measured complexity is 0 for every function
(for either choice of $M$).

As in Table~1,
the matrix measures $M_{\nz}$ and $M_{\abs}$ measure (respectively)
the maximum number of nonzero entries,
or the maximum sum of absolute values of entries,
in any pattern matrix row. (Their necessary properties are proven in Section~\ref{S:Definitions}.)

The kinds of actual complexity shown
are $C_{B_2}$ (all 0- or 1-input gates are free; all 2-input 1-output gates have cost 1) \cite{Wil11}\footnote{
    Our definition of $C_{B_2}$ differs from that reference in a trivial way,
    which is to give isolated NOT gates 0 cost. This makes no difference
    except for a function output which is the logical negation of an input variable,
    since all other NOT gates can be absorbed into connected gates without increasing their cost.
    We do this to simplify the behavior of cost when circuits are composed.
},
and ``free XORs'' or ``multiplicative complexity'' 
(the same, except $\F_2$-affine gates (like XOR) are also free) \cite{BPP00}.

\bigskip
About the measured complexity ($C_{M,\P}$) values themselves, we note:
\begin{itemize}

\item
The random pattern basis $\P_{\randOne}$ sees similarly high ``complexity'' in every function
(given its number of inputs),
as predicted in Sections~\ref{S:oftentrivial} and~\ref{S:evenfixedwidth};
the only exceptions are identity functions (not shown in the table), 
such as those computed by some circuits consisting only of wires.

\item
The monomial basis $\P_{\mon}$ (when used with $M_{\nz}$)
is measuring ``amount of lack of invertibility'' (which is also subadditive),
rather than actual complexity.
(The fact that it does this was mentioned in Section~\ref{S:evenfixedwidth}, and is easy to prove.
When used with $M_{\abs}$ instead, 
$\P_{\mon}$ also sees positive ``measured complexity'' 
in invertible functions which are not onto.)

\item
The ``candidate useful'' basis $\P_{\RmTwob}$ agrees with the ``free XORs'' kind of actual complexity (up to a constant factor). (This is not evidence that the ``free XORs'' kind of complexity is more natural~--- 
rather, it reflects the author's preexisting guess that it is,
since measuring $\F_2$-affine functions as having 0 complexity
was built into the definition of ``niceness'' which guided the search for that basis.)

\item
There is not yet any evidence about whether $M_{\nz}$ or $M_{\abs}$ is ``better''.\footnote{
    On the other hand, we can rule out using an $M$ 
    defined as the maximum {\it euclidean} norm of any pattern matrix row,
    since in any orthonormal pattern basis
    it would see all reversible functions as having 0 complexity.
}

\end{itemize}

\mydefs 

\section{Motivation\label{S:Motivation}}       

\mymotive

\section{Discussion\label{S:Discussion}}       

\mydiscuss


\section*{Acknowledgements}       

\addcontentsline{toc}{section}{Acknowledgements}

This paper wouldn't exist
without the excellent expository blogs
of Scott Aaronson and Timothy Gowers,
which introduced me to the status of circuit complexity
and to the ``natural proofs barrier''.
More generally,
I'm grateful to many authors for making readily available online their blog posts,
preprints, course notes, and Wikipedia edits;
as an amateur mathematician,
it would not otherwise be practical to become informed about new fields.

I also thank Scott Aaronson for valuable feedback and suggestions, 
and John Baez for useful discussions
and helping clarify some of my notation and terminology.
(Any errors are of course my own.)

\addcontentsline{toc}{section}{References} 


\input{patbasis.bbl}      


\end{document}

%% file: pbcircuit-modeleg.tex
\begin{tikzpicture}[scale=2.54]
\ifx\dpiclw\undefined\newdimen\dpiclw\fi
\global\def\dpicdraw{\draw[line width=\dpiclw]}
\global\def\dpicstop{;}
\dpiclw=0.8bp
\dpiclw=0.8bp
\dpicdraw (0.15,0.1875)
 --(0.35,0.1875)
 --(0.35,-0.1125)
 --(0.15,-0.1125)\dpicstop
\dpicdraw (0.15,-0.1125)
 ..controls (-0.05,-0.1125) and (-0.05,0.1875)
 ..(0.15,0.1875)\dpicstop
\dpicdraw (0.35,0.1125)
 --(0.5,0.1125)\dpicstop
\dpicdraw (0.5,0.1125)
 --(0.8,0.5625)\dpicstop
\dpicdraw (0.35,-0.0375)
 --(1.025,-0.0375)\dpicstop
\draw (1.025,-0.0375) node[right=-1.5bp]{$z$};
\dpicdraw (0.15,0.6375)
 --(0.35,0.6375)
 --(0.35,0.3375)
 --(0.15,0.3375)\dpicstop
\dpicdraw (0.15,0.3375)
 ..controls (-0.05,0.3375) and (-0.05,0.6375)
 ..(0.15,0.6375)\dpicstop
\dpicdraw (0.3875,0.5625) circle (0.014764in)\dpicstop
\dpicdraw (0.425,0.5625)
 --(1.025,0.5625)\dpicstop
\draw (1.025,0.5625) node[right=-1.5bp]{$x$};
\dpicdraw (0.35,0.4125)
 --(1.025,0.4125)\dpicstop
\draw (1.025,0.4125) node[right=-1.5bp]{$y$};
\dpicdraw (-0.366667,0.113589)
 --(-0.244444,0.113589)\dpicstop
\dpicdraw (-0.248957,0.111411)
 ..controls (-0.295009,0.20679) and (-0.295392,0.317894)
 ..(-0.25,0.413589)\dpicstop
\dpicdraw (-0.244444,0.413589)
 --(-0.366667,0.413589)\dpicstop
\dpicdraw (-0.366667,0.113589)
 ..controls (-0.467105,0.113589) and (-0.558302,0.172216)
 ..(-0.6,0.263589)\dpicstop
\dpicdraw (-0.366667,0.413589)
 ..controls (-0.467105,0.413589) and (-0.558302,0.354963)
 ..(-0.6,0.263589)\dpicstop
\filldraw[line width=0bp](-0.7,0.288589)
 --(-0.8,0.263589)
 --(-0.7,0.238589) --cycle\dpicstop
\dpicdraw (-0.6,0.263589)
 --(-0.777094,0.263589)\dpicstop
\dpicdraw (-0.275642,0.338589)
 --(0,0.4875)\dpicstop
\dpicdraw (-0.275642,0.188589)
 --(0,0.0375)\dpicstop
\end{tikzpicture}

%% file: pbcircuit-modeleg-comp.tex
\begin{tikzpicture}[scale=2.54]
\ifx\dpiclw\undefined\newdimen\dpiclw\fi
\global\def\dpicdraw{\draw[line width=\dpiclw]}
\global\def\dpicstop{;}
\dpiclw=0.8bp
\dpiclw=0.8bp
\dpicdraw (0.75,0.3375)
 --(0.95,0.3375)
 --(0.95,0.0375)
 --(0.75,0.0375)\dpicstop
\dpicdraw (0.75,0.0375)
 ..controls (0.55,0.0375) and (0.55,0.3375)
 ..(0.75,0.3375)\dpicstop
\dpicdraw (0.95,0.2625)
 --(1.25,0.2625)\dpicstop
\dpicdraw (1.25,0.2625)
 --(1.45,0.5625)\dpicstop
\dpicdraw (1.45,0.5625)
 --(2.375,0.5625)\dpicstop
\draw (2.375,0.5625) node[right=-1.5bp]{$x$};
\dpicdraw (0.95,0.1125)
 --(2.375,0.1125)\dpicstop
\draw (2.375,0.1125) node[right=-1.5bp]{$z$};
\dpicdraw[dotted](0.5,-0.4125)
 --(0.5,0.7875)\dpicstop
\dpicdraw[dotted](-0.1,-0.4125)
 --(-0.1,0.7875)\dpicstop
\dpicdraw[dotted](-0.7,-0.4125)
 --(-0.7,0.7875)\dpicstop
\dpicdraw[dotted](-1.3,-0.4125)
 --(-1.3,0.7875)\dpicstop
\dpicdraw[dotted](1.1,-0.4125)
 --(1.1,0.7875)\dpicstop
\dpicdraw[dotted](1.7,-0.4125)
 --(1.7,0.7875)\dpicstop
\dpicdraw[dotted](2.3,-0.4125)
 --(2.3,0.7875)\dpicstop
\dpicdraw (-0.45,0.6375)
 --(-0.25,0.6375)
 --(-0.25,0.3375)
 --(-0.45,0.3375)\dpicstop
\dpicdraw (-0.45,0.3375)
 ..controls (-0.65,0.3375) and (-0.65,0.6375)
 ..(-0.45,0.6375)\dpicstop
\dpicdraw (-0.2125,0.5625) circle (0.014764in)\dpicstop
\dpicdraw (-0.175,0.5625)
 --(0.025,0.5625)\dpicstop
\dpicdraw (0.025,0.5625)
 --(0.425,-0.1125)\dpicstop
\dpicdraw (0.425,-0.1125)
 --(1.15,-0.1125)\dpicstop
\dpicdraw (1.15,-0.1125)
 --(1.5,0.4125)\dpicstop
\dpicdraw (1.5,0.4125)
 --(1.95,0.4125)\dpicstop
\dpicdraw (1.95,0.4125)
 --(2.05,0.5625)\dpicstop
\dpicdraw (-0.25,0.4125)
 --(-0.025,0.4125)\dpicstop
\dpicdraw (-0.025,0.4125)
 --(0.375,-0.2625)\dpicstop
\dpicdraw (0.375,-0.2625)
 --(1.2,-0.2625)\dpicstop
\dpicdraw (1.2,-0.2625)
 --(1.55,0.2625)\dpicstop
\dpicdraw (1.55,0.2625)
 --(2.375,0.2625)\dpicstop
\draw (2.375,0.2625) node[right=-1.5bp]{$y$};
\dpicdraw (-0.966667,0.113589)
 --(-0.844444,0.113589)\dpicstop
\dpicdraw (-0.848957,0.111411)
 ..controls (-0.895009,0.20679) and (-0.895392,0.317894)
 ..(-0.85,0.413589)\dpicstop
\dpicdraw (-0.844444,0.413589)
 --(-0.966667,0.413589)\dpicstop
\dpicdraw (-0.966667,0.113589)
 ..controls (-1.067105,0.113589) and (-1.158302,0.172216)
 ..(-1.2,0.263589)\dpicstop
\dpicdraw (-0.966667,0.413589)
 ..controls (-1.067105,0.413589) and (-1.158302,0.354963)
 ..(-1.2,0.263589)\dpicstop
\filldraw[line width=0bp](-1.375,0.288589)
 --(-1.475,0.263589)
 --(-1.375,0.238589) --cycle\dpicstop
\dpicdraw (-1.2,0.263589)
 --(-1.452094,0.263589)\dpicstop
\dpicdraw (-0.875642,0.338589)
 --(-0.6,0.4875)\dpicstop
\dpicdraw (-0.875642,0.188589)
 --(0.6,0.1875)\dpicstop
\end{tikzpicture}

%% file: pbcircuit-NotXAndY.tex
\begin{tikzpicture}[scale=2.54,baseline]
\ifx\dpiclw\undefined\newdimen\dpiclw\fi
\global\def\dpicdraw{\draw[line width=\dpiclw]}
\global\def\dpicstop{;}
\dpiclw=0.8bp
\dpiclw=0.8bp
\dpicdraw (0.15,0.1875)
 --(0.35,0.1875)
 --(0.35,-0.1125)
 --(0.15,-0.1125)\dpicstop
\dpicdraw (0.15,-0.1125)
 ..controls (-0.05,-0.1125) and (-0.05,0.1875)
 ..(0.15,0.1875)\dpicstop
\dpicdraw (0.3875,0.1125) circle (0.014764in)\dpicstop
\filldraw[line width=0bp](-0.1,0.0625)
 --(-0.2,0.0375)
 --(-0.1,0.0125) --cycle\dpicstop
\dpicdraw (0,0.0375)
 --(-0.177094,0.0375)\dpicstop
\dpicdraw (0.425,0.1125)
 --(0.575,0.1125)\dpicstop
\draw (0.575,0.1125) node[right=-1.5bp]{$x$};
\dpicdraw (0.35,-0.0375)
 --(0.575,-0.0375)\dpicstop
\draw (0.575,-0.0375) node[right=-1.5bp]{$y$};
\end{tikzpicture}

%% file: pbcircuit-XAndY.tex
\begin{tikzpicture}[scale=2.54,baseline]
\ifx\dpiclw\undefined\newdimen\dpiclw\fi
\global\def\dpicdraw{\draw[line width=\dpiclw]}
\global\def\dpicstop{;}
\dpiclw=0.8bp
\dpiclw=0.8bp
\dpicdraw (0.15,0.1875)
 --(0.35,0.1875)
 --(0.35,-0.1125)
 --(0.15,-0.1125)\dpicstop
\dpicdraw (0.15,-0.1125)
 ..controls (-0.05,-0.1125) and (-0.05,0.1875)
 ..(0.15,0.1875)\dpicstop
\filldraw[line width=0bp](-0.1,0.0625)
 --(-0.2,0.0375)
 --(-0.1,0.0125) --cycle\dpicstop
\dpicdraw (0,0.0375)
 --(-0.177094,0.0375)\dpicstop
\dpicdraw (0.35,0.1125)
 --(0.575,0.1125)\dpicstop
\draw (0.575,0.1125) node[right=-1.5bp]{$x$};
\dpicdraw (0.35,-0.0375)
 --(0.575,-0.0375)\dpicstop
\draw (0.575,-0.0375) node[right=-1.5bp]{$y$};
\end{tikzpicture}

%% file: pbcircuit-NotWire.tex
\begin{tikzpicture}[scale=2.54,baseline]
\ifx\dpiclw\undefined\newdimen\dpiclw\fi
\global\def\dpicdraw{\draw[line width=\dpiclw]}
\global\def\dpicstop{;}
\dpiclw=0.8bp
\dpiclw=0.8bp
\filldraw[line width=0bp](0.675,0.1375)
 --(0.575,0.1125)
 --(0.675,0.0875) --cycle\dpicstop
\dpicdraw (0.597906,0.1125)
 --(0.775,0.1125)\dpicstop
\dpicdraw (0.8125,0.1125) circle (0.014764in)\dpicstop
\dpicdraw (0.85,0.1125)
 --(1,0.1125)\dpicstop
\draw (1,0.1125) node[right=-1.5bp]{$x$};
\filldraw[line width=0bp](0.675,-0.0125)
 --(0.575,-0.0375)
 --(0.675,-0.0625) --cycle\dpicstop
\dpicdraw (0.597906,-0.0375)
 --(1,-0.0375)\dpicstop
\draw (1,-0.0375) node[right=-1.5bp]{$y$};
\end{tikzpicture}

%% file: patbasis.bbl
\begin{thebibliography}{99}
\bibitem[--]{ref1}
        (Note: references to blog comments are labelled with my own summaries.)
\end{thebibliography}
